\documentclass[twocolumn,letterpaper,aps,prd,superscriptaddress,showpacs,nofootinbib,floatfix]{revtex4}

\usepackage{graphicx}	
\usepackage{xspace}     
\usepackage{subfigure}

\newenvironment{packed_enum}{
\begin{enumerate}
  \setlength{\itemsep}{1pt}
  \setlength{\parskip}{0pt}
  \setlength{\parsep}{0pt}
}{\end{enumerate}}

\newenvironment{packed_descr}{
\begin{description}
  \setlength{\itemsep}{1pt}
  \setlength{\parskip}{0pt}
  \setlength{\parsep}{0pt}
}{\end{description}}

\newcommand{\simlt}{\lower.5ex\hbox{$\; \buildrel < \over \sim \;$}}
\newcommand{\simgt}{\lower.5ex\hbox{$\; \buildrel > \over \sim \;$}}

\def\xmean{-1.6}
\def\xrmsp{0.5}
\def\xrmsn{0.4}

\def\dggsqrerrlim{3.3}
\def\dggsqrthreeerrlim{10.9}

\def\diffdggsqrerrmassp{0.4}
\def\diffdggsqrerrmassn{0.3}
\def\diffdggsqrerrscalep{0.7}
\def\diffdggsqrerrscalen{0.4}

\def\dggdssv{-0.006}
\def\dgggrsvs{0.016}
\def\dgggrsvv{0.019}

\begin{document}


\title{Double Spin Asymmetry of Electrons from Heavy Flavor Decays\\
in $p$$+$$p$ Collisions at $\sqrt{s}=200$ GeV}

\newcommand{\abilene}{Abilene Christian University, Abilene, Texas 79699, USA}
\newcommand{\banaras}{Department of Physics, Banaras Hindu University, Varanasi 221005, India}
\newcommand{\barc}{Bhabha Atomic Research Centre, Bombay 400 085, India}
\newcommand{\bnlcoll}{Collider-Accelerator Department, Brookhaven National Laboratory, Upton, New York 11973-5000, USA}
\newcommand{\bnlphys}{Physics Department, Brookhaven National Laboratory, Upton, New York 11973-5000, USA}
\newcommand{\caucr}{University of California - Riverside, Riverside, California 92521, USA}
\newcommand{\charlesczech}{Charles University, Ovocn\'{y} trh 5, Praha 1, 116 36, Prague, Czech Republic}
\newcommand{\chonbuk}{Chonbuk National University, Jeonju, 561-756, Korea}
\newcommand{\ciae}{Science and Technology on Nuclear Data Laboratory, China Institute of Atomic Energy, Beijing 102413, P.~R.~China}
\newcommand{\cns}{Center for Nuclear Study, Graduate School of Science, University of Tokyo, 7-3-1 Hongo, Bunkyo, Tokyo 113-0033, Japan}
\newcommand{\colorado}{University of Colorado, Boulder, Colorado 80309, USA}
\newcommand{\columbia}{Columbia University, New York, New York 10027 and Nevis Laboratories, Irvington, New York 10533, USA}
\newcommand{\czechtech}{Czech Technical University, Zikova 4, 166 36 Prague 6, Czech Republic}
\newcommand{\dapnia}{Dapnia, CEA Saclay, F-91191, Gif-sur-Yvette, France}
\newcommand{\elte}{ELTE, E{\"o}tv{\"o}s Lor{\'a}nd University, H - 1117 Budapest, P{\'a}zm{\'a}ny P. s. 1/A, Hungary}
\newcommand{\ewha}{Ewha Womans University, Seoul 120-750, Korea}
\newcommand{\fsu}{Florida State University, Tallahassee, Florida 32306, USA}
\newcommand{\gsu}{Georgia State University, Atlanta, Georgia 30303, USA}
\newcommand{\hanyang}{Hanyang University, Seoul 133-792, Korea}
\newcommand{\hiroshima}{Hiroshima University, Kagamiyama, Higashi-Hiroshima 739-8526, Japan}
\newcommand{\ihepprot}{IHEP Protvino, State Research Center of Russian Federation, Institute for High Energy Physics, Protvino, 142281, Russia}
\newcommand{\illuiuc}{University of Illinois at Urbana-Champaign, Urbana, Illinois 61801, USA}
\newcommand{\inrras}{Institute for Nuclear Research of the Russian Academy of Sciences, prospekt 60-letiya Oktyabrya 7a, Moscow 117312, Russia}
\newcommand{\instpasczech}{Institute of Physics, Academy of Sciences of the Czech Republic, Na Slovance 2, 182 21 Prague 8, Czech Republic}
\newcommand{\isu}{Iowa State University, Ames, Iowa 50011, USA}
\newcommand{\jinrdubna}{Joint Institute for Nuclear Research, 141980 Dubna, Moscow Region, Russia}
\newcommand{\jyvaskyla}{Helsinki Institute of Physics and University of Jyv{\"a}skyl{\"a}, P.O.Box 35, FI-40014 Jyv{\"a}skyl{\"a}, Finland}
\newcommand{\kek}{KEK, High Energy Accelerator Research Organization, Tsukuba, Ibaraki 305-0801, Japan}
\newcommand{\korea}{Korea University, Seoul, 136-701, Korea}
\newcommand{\kurchatov}{Russian Research Center ``Kurchatov Institute", Moscow, 123098 Russia}
\newcommand{\kyoto}{Kyoto University, Kyoto 606-8502, Japan}
\newcommand{\labllr}{Laboratoire Leprince-Ringuet, Ecole Polytechnique, CNRS-IN2P3, Route de Saclay, F-91128, Palaiseau, France}
\newcommand{\lawllnl}{Lawrence Livermore National Laboratory, Livermore, California 94550, USA}
\newcommand{\losalamos}{Los Alamos National Laboratory, Los Alamos, New Mexico 87545, USA}
\newcommand{\lpc}{LPC, Universit{\'e} Blaise Pascal, CNRS-IN2P3, Clermont-Fd, 63177 Aubiere Cedex, France}
\newcommand{\lund}{Department of Physics, Lund University, Box 118, SE-221 00 Lund, Sweden}
\newcommand{\maryland}{University of Maryland, College Park, Maryland 20742, USA}
\newcommand{\mass}{Department of Physics, University of Massachusetts, Amherst, Massachusetts 01003-9337, USA }
\newcommand{\muenster}{Institut fur Kernphysik, University of Muenster, D-48149 Muenster, Germany}
\newcommand{\muhlenberg}{Muhlenberg College, Allentown, Pennsylvania 18104-5586, USA}
\newcommand{\myongji}{Myongji University, Yongin, Kyonggido 449-728, Korea}
\newcommand{\nagasaki}{Nagasaki Institute of Applied Science, Nagasaki-shi, Nagasaki 851-0193, Japan}
\newcommand{\newmex}{University of New Mexico, Albuquerque, New Mexico 87131, USA }
\newcommand{\nmsu}{New Mexico State University, Las Cruces, New Mexico 88003, USA}
\newcommand{\ohio}{Department of Physics and Astronomy, Ohio University, Athens, Ohio 45701, USA}
\newcommand{\ornl}{Oak Ridge National Laboratory, Oak Ridge, Tennessee 37831, USA}
\newcommand{\orsay}{IPN-Orsay, Universite Paris Sud, CNRS-IN2P3, BP1, F-91406, Orsay, France}
\newcommand{\peking}{Peking University, Beijing 100871, P.~R.~China}
\newcommand{\pnpi}{PNPI, Petersburg Nuclear Physics Institute, Gatchina, Leningrad region, 188300, Russia}
\newcommand{\riken}{RIKEN Nishina Center for Accelerator-Based Science, Wako, Saitama 351-0198, Japan}
\newcommand{\rikjrbrc}{RIKEN BNL Research Center, Brookhaven National Laboratory, Upton, New York 11973-5000, USA}
\newcommand{\rikkyo}{Physics Department, Rikkyo University, 3-34-1 Nishi-Ikebukuro, Toshima, Tokyo 171-8501, Japan}
\newcommand{\saispbstu}{Saint Petersburg State Polytechnic University, St. Petersburg, 195251 Russia}
\newcommand{\saopaulo}{Universidade de S{\~a}o Paulo, Instituto de F\'{\i}sica, Caixa Postal 66318, S{\~a}o Paulo CEP05315-970, Brazil}
\newcommand{\seoulnat}{Seoul National University, Seoul, Korea}
\newcommand{\stonybrkc}{Chemistry Department, Stony Brook University, SUNY, Stony Brook, New York 11794-3400, USA}
\newcommand{\stonycrkp}{Department of Physics and Astronomy, Stony Brook University, SUNY, Stony Brook, New York 11794-3400, USA}
\newcommand{\tenn}{University of Tennessee, Knoxville, Tennessee 37996, USA}
\newcommand{\titech}{Department of Physics, Tokyo Institute of Technology, Oh-okayama, Meguro, Tokyo 152-8551, Japan}
\newcommand{\tsukuba}{Institute of Physics, University of Tsukuba, Tsukuba, Ibaraki 305, Japan}
\newcommand{\vandy}{Vanderbilt University, Nashville, Tennessee 37235, USA}
\newcommand{\weizmann}{Weizmann Institute, Rehovot 76100, Israel}
\newcommand{\wigner}{Institute for Particle and Nuclear Physics, Wigner Research Centre for Physics, Hungarian Academy of Sciences (Wigner RCP, RMKI) H-1525 Budapest 114, POBox 49, Budapest, Hungary}
\newcommand{\yonsei}{Yonsei University, IPAP, Seoul 120-749, Korea}
\affiliation{\abilene}
\affiliation{\banaras}
\affiliation{\barc}
\affiliation{\bnlcoll}
\affiliation{\bnlphys}
\affiliation{\caucr}
\affiliation{\charlesczech}
\affiliation{\chonbuk}
\affiliation{\ciae}
\affiliation{\cns}
\affiliation{\colorado}
\affiliation{\columbia}
\affiliation{\czechtech}
\affiliation{\dapnia}
\affiliation{\elte}
\affiliation{\ewha}
\affiliation{\fsu}
\affiliation{\gsu}
\affiliation{\hanyang}
\affiliation{\hiroshima}
\affiliation{\ihepprot}
\affiliation{\illuiuc}
\affiliation{\inrras}
\affiliation{\instpasczech}
\affiliation{\isu}
\affiliation{\jinrdubna}
\affiliation{\jyvaskyla}
\affiliation{\kek}
\affiliation{\korea}
\affiliation{\kurchatov}
\affiliation{\kyoto}
\affiliation{\labllr}
\affiliation{\lawllnl}
\affiliation{\losalamos}
\affiliation{\lpc}
\affiliation{\lund}
\affiliation{\maryland}
\affiliation{\mass}
\affiliation{\muenster}
\affiliation{\muhlenberg}
\affiliation{\myongji}
\affiliation{\nagasaki}
\affiliation{\newmex}
\affiliation{\nmsu}
\affiliation{\ohio}
\affiliation{\ornl}
\affiliation{\orsay}
\affiliation{\peking}
\affiliation{\pnpi}
\affiliation{\riken}
\affiliation{\rikjrbrc}
\affiliation{\rikkyo}
\affiliation{\saispbstu}
\affiliation{\saopaulo}
\affiliation{\seoulnat}
\affiliation{\stonybrkc}
\affiliation{\stonycrkp}
\affiliation{\tenn}
\affiliation{\titech}
\affiliation{\tsukuba}
\affiliation{\vandy}
\affiliation{\weizmann}
\affiliation{\wigner}
\affiliation{\yonsei}
\author{A.~Adare} \affiliation{\colorado}
\author{S.~Afanasiev} \affiliation{\jinrdubna}
\author{C.~Aidala} \affiliation{\losalamos}
\author{N.N.~Ajitanand} \affiliation{\stonybrkc}
\author{Y.~Akiba} \affiliation{\riken} \affiliation{\rikjrbrc}
\author{R.~Akimoto} \affiliation{\cns}
\author{H.~Al-Ta'ani} \affiliation{\nmsu}
\author{J.~Alexander} \affiliation{\stonybrkc}
\author{K.R.~Andrews} \affiliation{\abilene}
\author{A.~Angerami} \affiliation{\columbia}
% ...
\collaboration{PHENIX Collaboration} \noaffiliation

\date{\today}

\begin{abstract}

We report on the first measurement of double-spin asymmetry, 
$A_{LL}$, of electrons from the decays of hadrons containing 
heavy flavor in longitudinally polarized $p$$+$$p$ collisions 
at $\sqrt{s}=200$ GeV for $p_T=$ 0.5 to 3.0~GeV/$c$. The 
asymmetry was measured at midrapidity ($|\eta|<0.35$) with the 
PHENIX detector at the Relativistic Heavy Ion Collider. The 
measured asymmetries are consistent with zero within the 
statistical errors. We obtained a constraint for the polarized 
gluon distribution in the proton of
$|\Delta g/g(\log_{10}x= -1.6^{+0.5}_{-0.4}, 
\mu=m_{T}^{c})|^{2}<0.033~(1\sigma$) based on a 
leading-order perturbative-quantum-chromodynamics model, 
using the measured asymmetry.

\end{abstract}

\pacs{13.85.Ni,13.88.+e,14.20.Dh,25.75.Dw}

\maketitle

\section{\label{sec:Introduction}Introduction}

The measurement of the first moment of the proton's spin-dependent structure function
$g_{1}^{p}$ by the European Muon Collaboration
(EMC) \cite{EMCg1p:PLB1988,EMCg1p:NPB1989}
revealed a discrepancy from the Ellis-Jaffe sum rule
\cite{EllisJaffe:PRD1974,EllisJaffeCorr:NPB1980} 
and also the fact that the SU(3) flavor-singlet axial
charge $g_{\rm{A}}^{(0)}$ was smaller than
expected from the static and
relativistic quark models \cite{ProtonSpin:2008}.
After these discoveries, experimental
efforts \cite{HERMESg1:PRD2007,COMPASSg1d:PLB2007:1,COMPASSg1d:PLB2007:2}
focused on a detailed understanding of the spin structure of the
proton.
The proton spin $s_z/\hbar=1/2$ can be decomposed as 
$\frac{1}{2}=\frac{1}{2}\Delta \Sigma + \Delta G + L_{z}$
from conservation of angular momentum.
The measurements precisely determined the total spin 
carried by quarks and anti-quarks, $\Delta \Sigma$, which is only
about $30$\% of the proton spin.
The remaining proton spin can be attributed to the other components, 
the gluon spin contribution ($\Delta G$) and/or orbital angular momentum contributions ($L_{z}$).
The total gluon polarization is given by
\begin{eqnarray}
 \label{eq:GluonPolarization}
\Delta G (\mu) \equiv \int_{0}^{1} dx \Delta g(x,\mu),
\end{eqnarray}
where $x$ and $\mu$ represent Bjorken $x$ and factorization scale
respectively.
The challenge for the $\Delta G(\mu)$ determination
is to precisely map the gluon polarization density $\Delta
g (x,\mu)$ over a wide range of $x$.

The Relativistic Heavy Ion Collider (RHIC),
which can accelerate polarized
proton beams up to 255~GeV,
is a unique and powerful
facility to study the gluon polarization.
One of the main goals of the RHIC physics program is to determine the gluon 
polarization through measurements of longitudinal double-spin asymmetries,

\begin{eqnarray}
 \label{eq:ALLDef}
A_{LL}\equiv\frac{\sigma^{++}-\sigma^{+-}}{\sigma^{++}+\sigma^{+-}},
\end{eqnarray}

\noindent where $\sigma^{++}$ and $\sigma^{+-}$ 
denote the cross sections of a specific process in the 
polarized $p$$+$$p$ collisions with same and opposite 
helicities. Using $A_{LL}$, the polarized cross sections, 
$\sigma^{++}$ and $\sigma^{+-}$, can be represented as,

\begin{eqnarray}
\label{eq:PolCrossSection}
 \sigma^{+\pm} = \sigma_{0}(1\pm A_{LL}),
\end{eqnarray}

\noindent where $\sigma_{0}$ is the unpolarized cross section 
of the
process. 
$A_{LL}$ has been measured previously in several channels by PHENIX and
STAR, including inclusive 
$\pi^{0}$ \cite{PHENIXPi0ALL:PRL2009,PHENIXPi0ALL:PRD2007,PHENIXPi0ALL:PRD2006,PHENIXPi0ALL:PRD2009},
$\eta$ \cite{PHENIXEtaALL:PRD2011}, and
jet \cite{STARJetALL:arXiv2011,STARJetALL:PRL2008,PHENIXJetALL:PRD2011}
production.

Using the measured asymmetries, as well as the world-data on polarized
inclusive and semi-inclusive deep-inelastic scattering
(DIS) \cite{HERMESg1:PRD2007,COMPASSg1d:PLB2007:1,COMPASSg1d:PLB2007:2,HERMESpPDF:PRD2005,COMPASSSIDIS:PLB2008},  
a global analysis based on perturbative-quantum-chromodynamics 
(pQCD) calculation was performed at
next-to-leading order (NLO) in the strong-coupling constant
$\alpha_{S}$~\cite{DSSV:PRD2009}. 
The resulting $\Delta g (x,\mu)$ from the best fit is
too small to explain the proton spin in the Bjorken $x$ range of
$0.05<x<0.2$ ($-1.3<\log_{10}x<-0.7$) without considering $L_{z}$,
though a substantial gluon polarization is not ruled out yet due to the
uncertainties. 
Also, due to the limited Bjorken $x$ coverage,
there is a sizable uncertainty in
Eq.~\ref{eq:GluonPolarization} from the unexplored small
$x$ region.

The polarized cross section of heavy flavor production
on the partonic level is well
studied with leading-order (LO) and NLO pQCD
calculations \cite{LOHFXSect:PLB1994,NLOHFXSect:PRD2003,NLOHFXSect:ArXiv2000}.
The heavy quarks are produced dominantly by the gluon-gluon
interaction at the partonic level \cite{HQSpin:arXiv2009}.
Therefore, this channel has good sensitivity to the polarized gluon
density.
In addition, the large mass of the heavy quark
ensures that pQCD techniques are applicable for calculations of the
cross section.
Therefore, the measurement of heavy flavor production in polarized
proton collisions is a useful tool to study gluon polarization. 

In $p$$+$$p$ collisions at $\sqrt{s}=200$~GeV, the heavy flavor
production below $p_{T}\sim5$~GeV/$c$
is dominated by charm quarks.  
The Bjorken $x$ region covered by this process at midrapidity
is centered around $2m_{c}/\sqrt{s}\sim1.4\times10^{-2}$ where $m_{c}$
represents the charm quark mass.
Hence, measurement of the spin dependent heavy flavor production
is sensitive to the unexplored $x$ region, and
complements other data on the total gluon polarization $\Delta G(\mu)$.

At PHENIX, hadrons containing heavy flavors are measured through their
semi-leptonic decays to electrons and
positrons (heavy flavor electrons) \cite{SingleE:PRL2006,HQPhenix:PRC2011}.
Therefore the double-spin asymmetry of the heavy flavor electrons is an
important measurement for the gluon polarization study.
In this paper, we report the first measurement of this asymmetry,
and a resulting constraint on the gluon polarization with an LO
pQCD calculation. 

The organization of this paper is as follows:
We introduce the PHENIX detector system used for the measurement in
Sec.~\ref{sec:ExperimentalSetup}.
The method for the heavy flavor electron analysis is discussed in
Sec.~\ref{sec:MeasurementOfSE} and the results of the cross section and the
spin asymmetry are shown in Sec.~\ref{sec:CrossSection} and
Sec.~\ref{sec:HelicityAsymmetry}, respectively.
From the asymmetry result, we estimate a constraint on the polarized
gluon density, which is described in Sec.~\ref{sec:Discussion}.
For the sake of simplicity,
we use the word ``electron'' to include both electron and positron
throughout this paper, and distinguish by charge where necessary.

 \section{\label{sec:ExperimentalSetup}Experimental Setup}

This measurement is performed with the PHENIX detector positioned at
one of collision points at RHIC.
The RHIC accelerator comprises the blue ring
circulating clockwise and the yellow ring
circulating counter-clockwise.
For this experiment,
polarized bunches are stored and accelerated up to 100~GeV in each ring
and collide with longitudinal polarizations of
$\sim57$\% along the beams at the collision point with a collision
energy of $\sqrt{s}=200$~GeV.
The bunch polarizations are changed to parallel (beam-helicity $+$)
or anti-parallel (beam-helicity $-$) along the beams
alternately in the collisions
to realize all 4 ($=2\times 2$) combinations of the crossing beam-helicities.
Each time the accelerator is filled, the pattern of beam helicities in
the bunches is changed, in order to confirm the absence of a pattern
dependence of the measured spin asymmetry.
See Sec.~\ref{sec:HelicityAsymmetry} for details.

A detailed description of the complete PHENIX detector system 
can be found elsewhere 
\cite{PHENIXOV:NIM2003,PHENIXmag:NIM2003,PHENIXcnt:NIM2003,PHENIXpid:NIM2003,PHENIXemc:NIM2003,PHENIXmuon:NIM2003,PHENIXin:NIM2003}. 
The main detectors that are used in this analysis are beam-beam 
counters (BBC), zero degree calorimeters (ZDC), and two central 
arm spectrometers. The BBC provides the collision vertex 
information and the minimum bias (MB) trigger. The luminosity 
is determined by the number of MB triggers. Electrons are 
measured with the two central spectrometer arms which each 
cover a pseudorapidity range of $|\eta|<0.35$ and azimuthal 
angle $\Delta\phi=\pi/2$. 

\begin{figure}[tbh]
    \includegraphics[width=1.0\linewidth,clip]{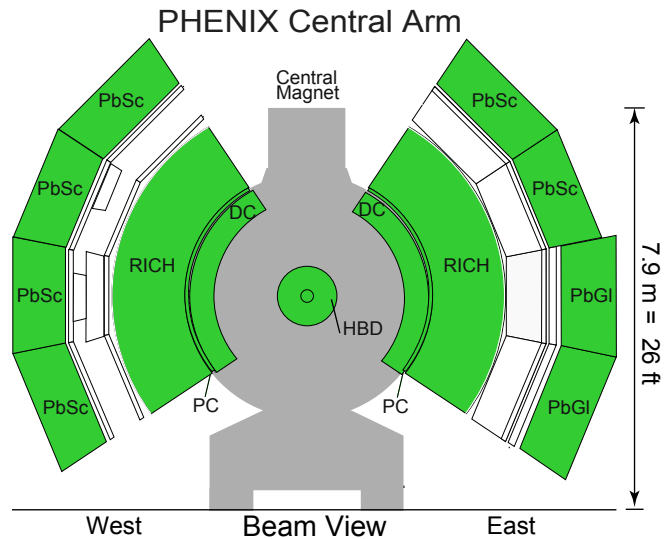}
\caption{\label{fig:PHENIXDetector}(color online)
Beam view (at $z=0$) of the PHENIX central arm detectors in 
2009. See text for details.}
\end{figure}

Figure~\ref{fig:PHENIXDetector} shows 
the beam view of the 2009 PHENIX central arms configuration, 
which comprises the central magnet (CM), drift chamber (DC), 
and pad chamber (PC) [for charged particle tracking], the 
ring-imaging \v{C}erenkov detector (RICH) and hadron blind 
detector (HBD)~\cite{HBD:NIM2004,HBD:NIM2011} [for electron 
identification], and the electromagnetic calorimeter (EMCal) 
[for energy measurement]. Below we summarize the features of 
the detectors and the CM.

The BBCs are two identical counters positioned at $\pm1.44$~m 
from the nominal interaction point along the beam direction and 
cover pseudorapidity of $3.1<|\eta|<3.9$. They measure the 
collision vertex along the beam axis by measuring the time 
difference between the two counters, and also provide the MB 
trigger defined by at least one hit on each side of the vertex. 
The position resolution for the vertex is $\sim2.0$~cm in 
$p$$+$$p$ collision.

The ZDCs, which are located at $\pm18.0$~m away from the 
nominal interaction point along the beam direction, detect 
neutral particles near the beam axis ($\theta<2.5$~mrad). Along 
with the BBCs, the trigger counts recorded by the ZDCs are used 
to determine the relative luminosity between crossings with 
different beam-helicities combinations. The ZDCs also serve for 
monitoring the orientation of the beam polarization in the 
PHENIX interaction region through the experiment.


The transverse momentum of a charged particle track
is determined by its
curvature in the magnetic field provided by the PHENIX CM
system \cite{PHENIXmag:NIM2003}. 
The CM is energized by two pairs of concentric coils and provides an
axial magnetic field parallel to the beam direction.
During this measurement, the two coils of the CM were operated in the
canceling (``$+-$'') configuration.
This configuration
is essential for the background rejection of the heavy flavor electron
measurement with the HBD as described later.
In this configuration, the field is almost
canceled out around the beam axis in the radial region $0<R<50$~cm, and has a peak value of
$\sim0.35$~T around $R\sim100$~cm.
The total field integral is $|\int B\times dl|=0.43$~Tm.


The DC and PC in the central arms measure charged particle trajectories
in the azimuthal direction to determine the transverse momentum
($p_{T}$) of each particle.
By combining the polar angle measured by
the PC and the vertex information along the beam axis from the BBC with
$p_{T}$, the total momentum $p$ is determined.
The DC is positioned between $202$~cm and $246$~cm in radial distance from
the collision point for both the west and east arms and the PC is
$247$-$252$~cm.


The RICH is a threshold-type gas $\check{\rm{C}}$erenkov counter and the
primary detector used to identify electrons in PHENIX.
It is located in the radial region of $2.5$-$4.1$~m.
The RICH has a $\check{\rm{C}}$erenkov threshold of $\gamma = 35$, which
corresponds to $p=20$~MeV/$c$ for electrons and $p=4.9$~GeV/$c$ for charged
pions.


The EMCal comprises four rectangular sectors in each arm. 
The six sectors based on lead-scintillator calorimetry and 
the two (lowest sectors on the east arm) based on lead-glass 
calorimetry are positioned at radial distances from the 
collision point of $\sim5.1$~m and $\sim5.4$~m, 
respectively.

\begin{figure}[tbh]
\subfigure[ HBD from top view.]
{\includegraphics[bb=0 0 591 553,width=1.0\linewidth]{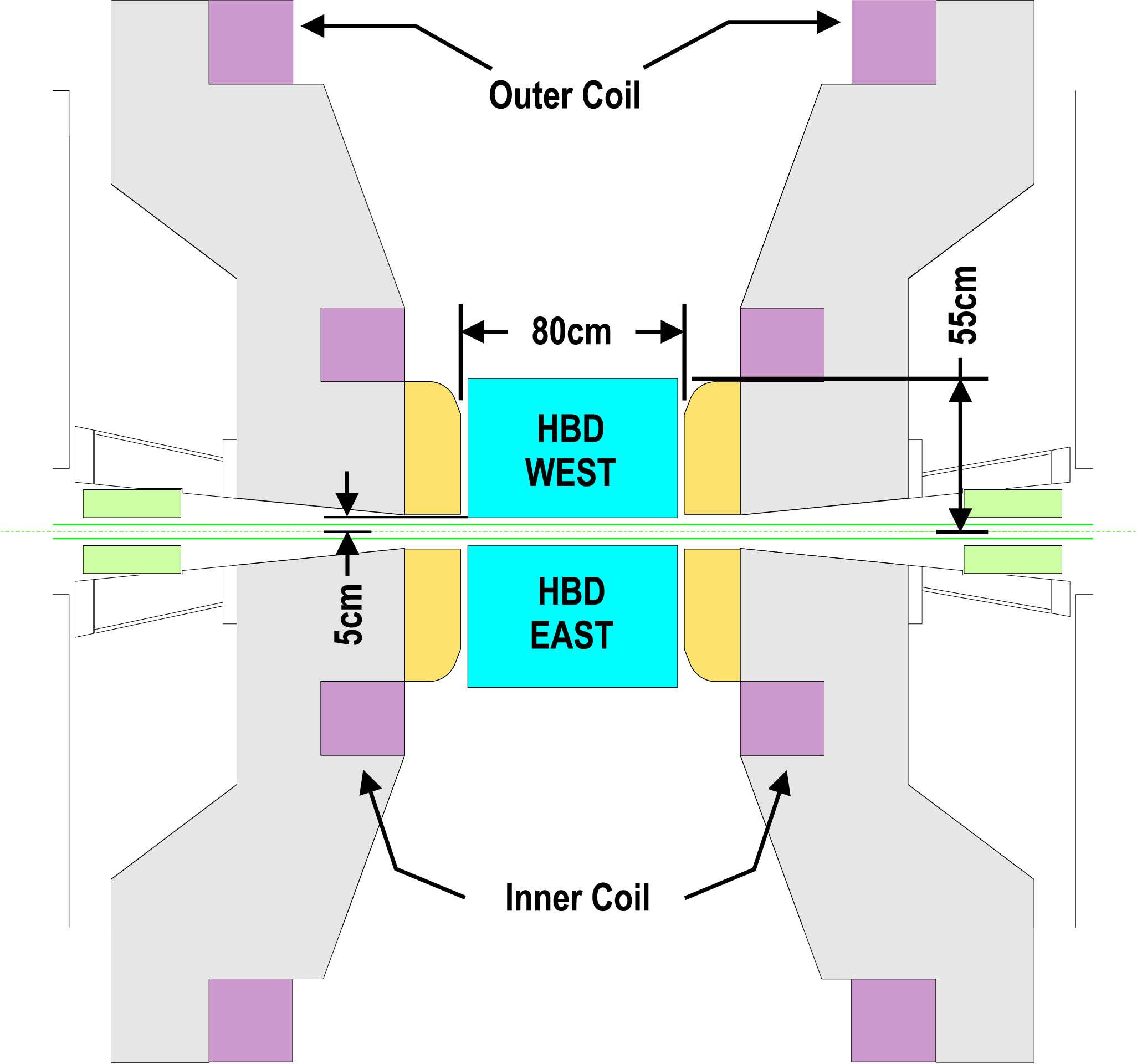}}
\subfigure[ HBD exploded view.]
{\includegraphics[width=1.0\linewidth,clip]{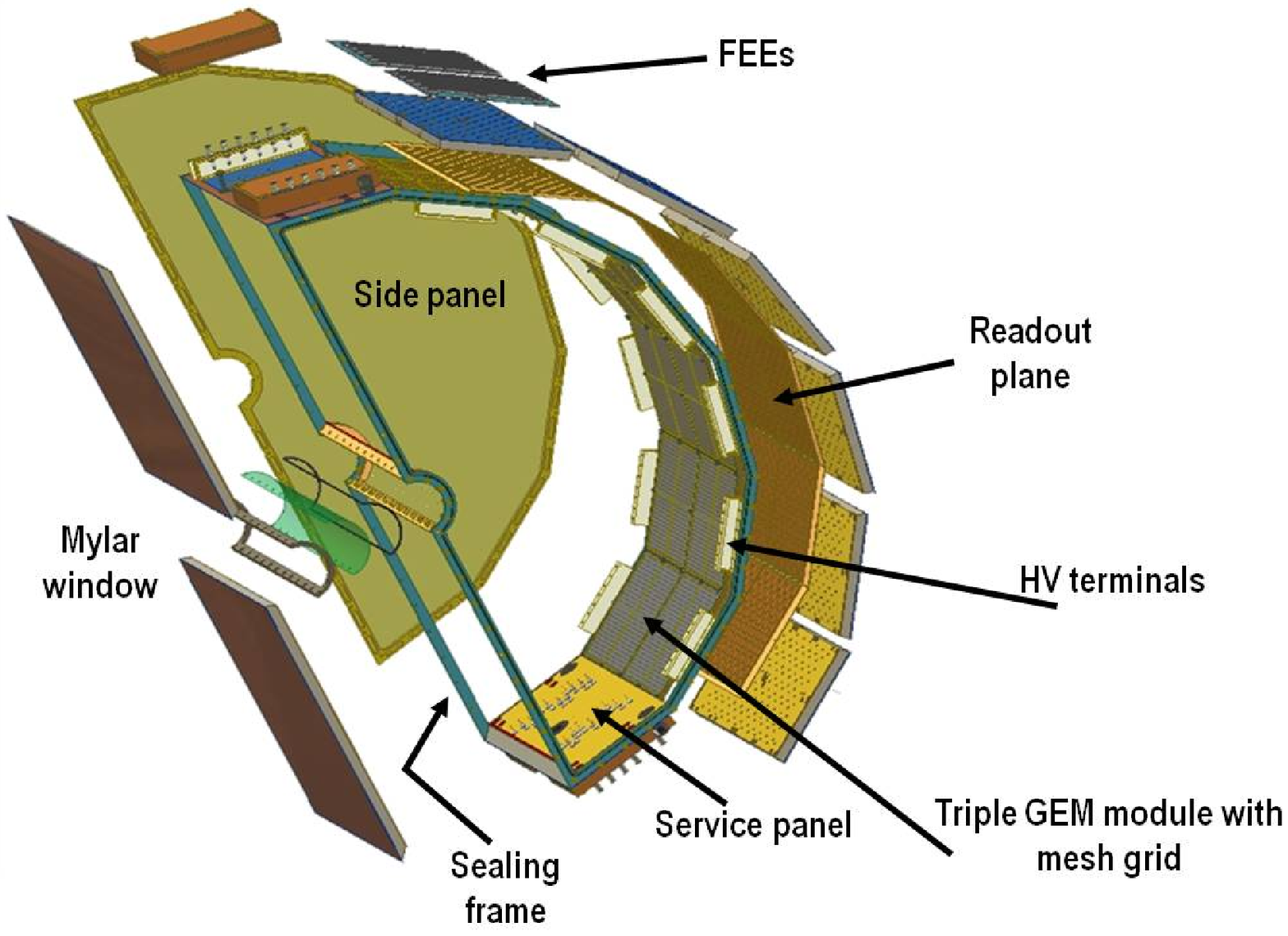}}
\caption{\label{fig:HBD}(color online) 
(a) Top view of the HBD showing the location of the HBD in the 
central magnet. (b) Exploded view of one half HBD arm. CF$_{4}$ 
gas is filled in the volume as the $\check{\rm{C}}$erenkov 
light radiator.
}
\end{figure}


A challenging issue for the heavy flavor electron 
measurement is to reject the dominant background of electron 
pairs from $\gamma$ conversions and Dalitz decays of 
$\pi^{0}$ and $\eta$ mesons, which are mediated by virtual 
photons. These electrons are called ``photonic electrons'', 
while all the other electrons are called ``nonphotonic 
electrons''. Most nonphotonic electrons are from heavy 
flavor decays, however, electrons from $K_{e3}$ decays 
($K\rightarrow e\nu \pi$) and the dielectron decays of light 
vector mesons are also nonphotonic~\cite{SingleE:PRL2006}. 
The HBD aims to considerably reduce the photonic electron 
pairs utilizing distinctive feature of the $e^{+}e^{-}$ 
pairs, namely their small opening angles.

The HBD is a position-sensitive $\check{\rm{C}}$erenkov detector
operated with pure CF$_{4}$ gas as a radiator.
It covers pseudorapidity $|\eta|<0.45$ and $2\times 3\pi /4$ in
azimuth.
The coverage is larger than the acceptance of the other
detectors in the central arm in order to detect 
photonic electron pairs with only one track reconstructed
in the central arm and the other outside of the central arm acceptance.
Figure~\ref{fig:HBD} shows the top view and exploded view of the HBD.
The HBD has a $50$~cm long radiator directly coupled in a
windowless configuration to a readout element consisting of a triple
Gas Electron Multiplier (GEM) stack,
with a CsI photocathode evaporated on the top surface of the
GEM facing the collision point
and pad readout at the exterior of the stack.
The readout element in each HBD arm is divided into five sectors.
The expected number of photoelectrons for an electron track is about 20,
which is consistent with the measured number.
Since the HBD is placed close to the collision point, the material
thickness is small in order to minimize conversions.
The total thickness to pass through the HBD
is $0.024X_{0}$ and the thickness before the GEM pads is
$0.007X_{0}$.

The $\check{\rm{C}}$erenkov light generated by electrons
is directly collected on a photosensitive cathode plane, forming
an almost circular blob image.
The readout pad plane comprises hexagonal pads with an area of 
$6.2$~cm$^2$ (hexagon side length $a=1.55$~cm) which is comparable to, but
smaller than, the blob size which has a maximum area of $9.9$~cm$^2$.

The HBD is located in a field free region that preserves the
original direction of the $e^{+}e^{-}$ pair.
The $\check{\rm{C}}$erenkov blobs created by electron pairs with a small
opening angle overlap, and therefore generate a signal in the HBD with
twice the amplitude of a single electron.
Electrons originating from $\pi^{0}$ and $\eta$ Dalitz decays and
$\gamma$ conversions can largely be eliminated by rejecting tracks which
correspond to large signals in the HBD.
 \section{\label{sec:MeasurementOfSE}Heavy Flavor Electron Analysis}

With the improved signal purity from the HBD,
the double helicity asymmetry of the
heavy flavor electrons was measured.
In this section, we explain how the heavy flavor electron analysis and the
purification of the heavy flavor electron sample using the HBD was performed.

\subsection{\label{subsec:DataSet}Data Set}

The data used here were recorded by PHENIX during 2009.
The data set was selected by a level-1 electron trigger in coincidence
with the MB trigger. 
The electron trigger required a minimum energy deposit of $0.6$~GeV in a
$2\times2$ tile of towers in EMCal,
$\check{\rm{C}}$erenkov light detection in the
RICH,
and acceptance matching of these two hits.
After a vertex cut of $|z_{{\rm vtx}}|<20$~cm and
data quality cuts, 
an equivalent of
$1.4\times10^{11}$ MB events, corresponding to $6.1$~pb$^{-1}$,
sampled by the electron trigger were analyzed.

\subsection{\label{subsec:ElectronSelection}Electron Selection}

Electrons are reconstructed using the detectors in the PHENIX 
central arm described above. Several useful variables for the 
electron selection which were used in the previous electron 
analysis in 2006 \cite{HQPhenix:PRC2011} are also used in this 
analysis. In addition to the conventional parameters, we 
introduced a new value, $q_{\rm{clus}}$, for the HBD analysis.

\begin{description}
 \item[hbdcharge: $q_{\rm{clus}}$]
Total charge of the associated HBD cluster calibrated 
in units of the number of photoelectrons (p.e.).
\end{description}

The electron selection cuts (eID-cut) are:

\begin{packed_enum}
\item[] 4.0$\sigma$ matching between track and EMCal cluster 
\item[]  \# of hit tubes in RICH around track $\geq2$ 
\item[]  3.5$\sigma$ matching between track and HBD cluster 
\item[]  shower profile cut on EMCal 
\item[]  $0.57<E/p<1.37$ ($0.5$ GeV/$c$ $<p_{T}<1.0$ GeV/$c$) 
\item[]  $0.60<E/p<1.32$ ($1.0$ GeV/$c$ $<p_{T}<1.5$ GeV/$c$) 
\item[]  $0.64<E/p<1.28$ ($1.5$ GeV/$c$ $<p_{T}<5.0$ GeV/$c$) 
\item[]  \# of hit pads in HBD cluster $\geq2$ 
\item[]  $q_{\rm{clus}}>8.0$ p.e. 
\item[]  ($q_{\rm{clus}}>4.0$ p.e. for one low-gain HBD sector) 
\end{packed_enum}

These cuts require hits in the HBD, RICH, and EMCal that are associated
with projections of the track onto these detectors.
The shower profile in the EMCal is required to match the profile
expected of an electromagnetic shower.
For electrons, the energy deposit on EMCal, $E$, and the
magnitude of the reconstructed momentum on DC and PC,
$p$, should match due to their small mass. 
Therefore the ratio, $E/p$, was required to be close to 1.
Since the energy resolution of the EMCal depends on the momentum of the
electron, the cut boundaries were changed in different momentum
range.
Charged particles traversing the CF$_4$ volume in the HBD produce
also scintillation light, which has no directivity and creates
hits with small charge in random locations in the GEM pads.
To remove HBD background hits 
by the scintillation light,
a minimum charge and a minimum cluster size
were required for the HBD hit clusters.
During this measurement, the efficiency for the $\check{\rm{C}}$erenkov
light in one HBD sector was low compared with other sectors.
Hence we apply a different charge cut to that HBD sector for the
electron selection.

\begin{figure}[tbh]
\includegraphics[width=1.0\linewidth,clip]{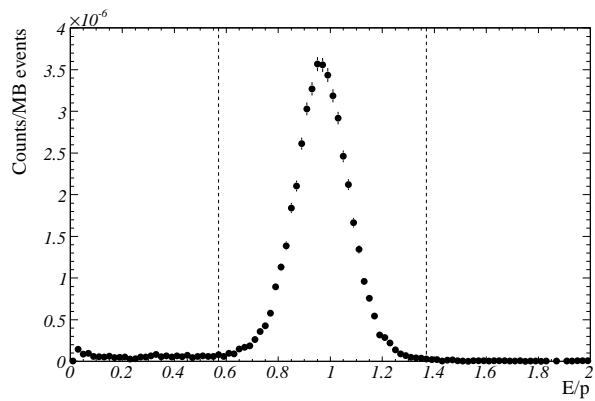}
\caption{\label{fig:EPRatio}$E/p$ distributions
for $0.5$~GeV/$c$~$<p_{T}<1.0$~GeV/$c$ reconstructed charged tracks
with the eID-Cut other than the $E/p$ cut.
Criteria of the $E/p$ cut for the momentum region
are shown by dashed lines in the plot.}
\end{figure}

The $E/p$ distribution for tracks selected with these cuts
is shown in Fig.~\ref{fig:EPRatio}.
The clear peak around $E/p=1$ corresponds to 
electrons and the spread of events around the peak consists mainly of
electrons from $K_{e3}$ decays and misidentified hadrons.
As the figure shows, the fraction of these background tracks in the
reconstructed electron sample after applying eID-Cut including the
$E/p$ cut was small.
The fractions of the $K_{e3}$ decays and the misidentified hadrons are 
described in Sec.~\ref{subsec:SystematicErrorInMSE} and
Sec.~\ref{subsec:SEResult}.

As mentioned in Sec.~\ref{sec:ExperimentalSetup}, we remove
the photonic electrons and purify the heavy flavor electrons on the
basis of the associated HBD cluster charge.
The nonphotonic electron cuts (npe-Cut) are:

\begin{packed_enum}
\item[] $8.0<q_{\rm{clus}}<28.0$ p.e.
\item[] ($4.0<q_{\rm{clus}}<17.0$ p.e. for 1 low-gain HBD sector)
\end{packed_enum}

\subsection{\label{subsec:SEYieldEstimation}Yield estimation of heavy flavor
electrons with HBD}

\begin{figure}[tbh]
\subfigure[\label{fig:HBDResponse:1} Non-photonic electrons.]
{\includegraphics[width=0.49\linewidth,clip]{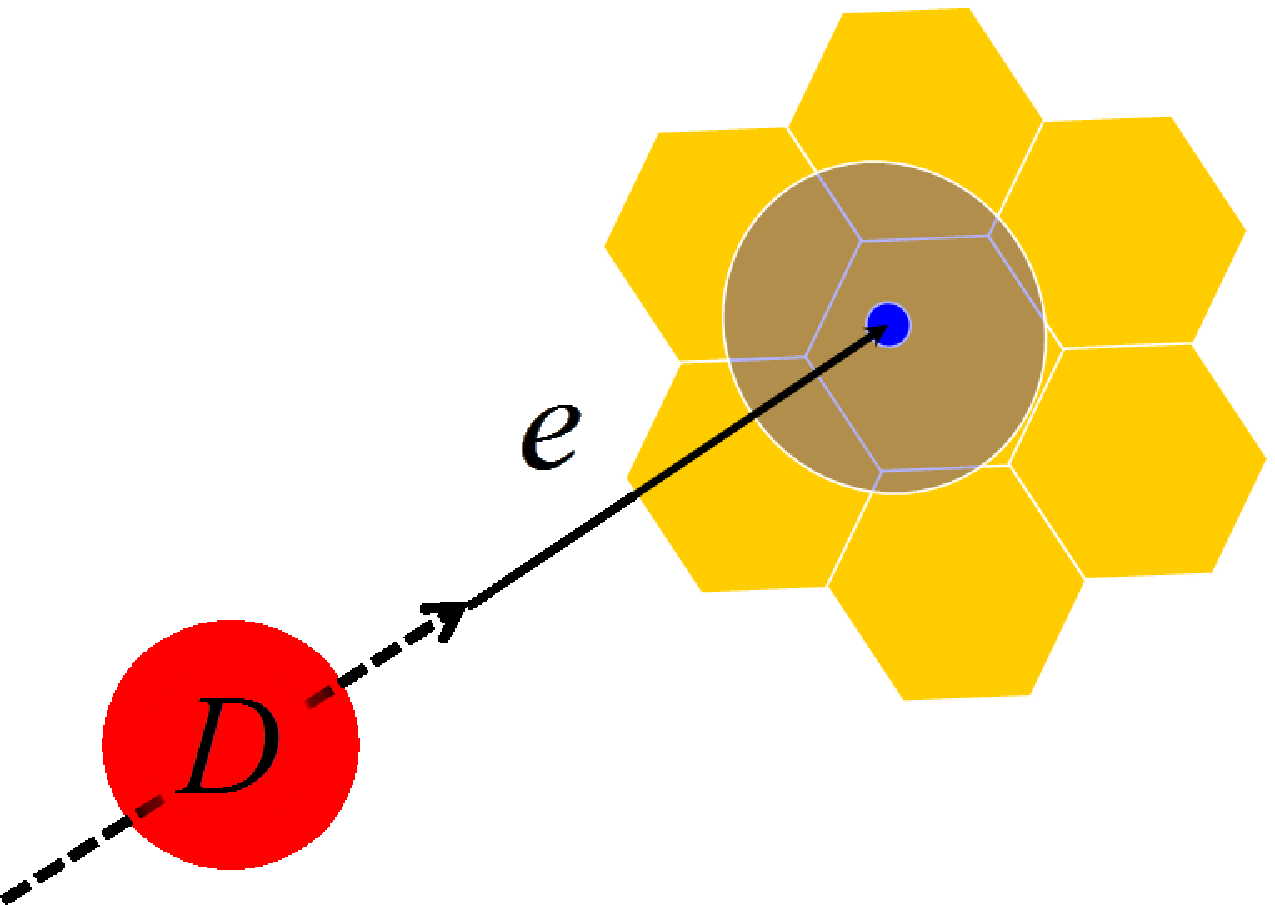}}\\
\subfigure[\label{fig:HBDResponse:2} Photonic electrons 
(merging cluster).]
{\includegraphics[width=0.49\linewidth,clip]{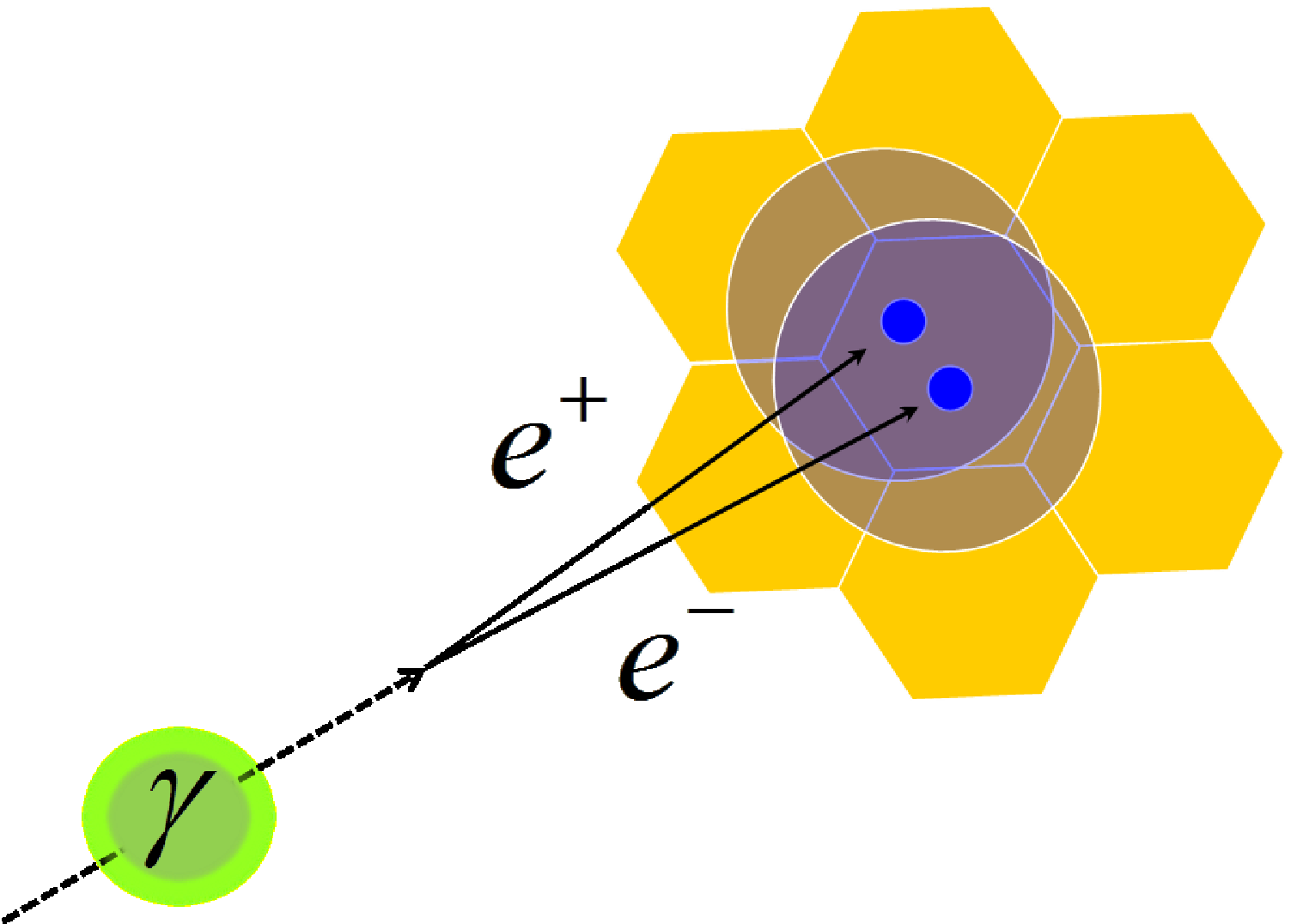}}
\subfigure[\label{fig:HBDResponse:3} Photonic electrons 
(separated clusters).]
{\includegraphics[width=0.49\linewidth,clip]{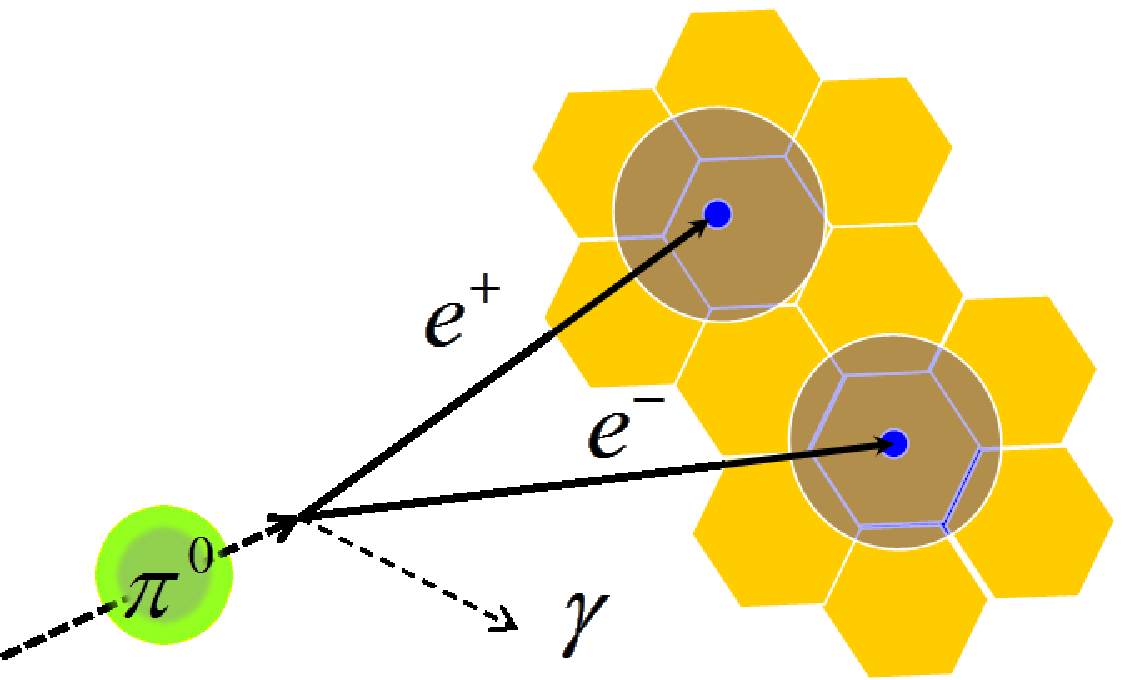}}
\caption{\label{fig:HBDResponse}(color online)
Responses of the HBD for (a) nonphotonic electrons and (b,c) 
photonic electrons. (b) Most of the photonic electron pair 
create merging clusters. (c) However, the photonic 
electrons with large opening angles create separated clusters.
}
\end{figure}

We categorize the HBD hit clusters into three types according to the
source of the cluster.
A cluster created by a single blob of $\rm{\check{C}}$erenkov light
from a nonphotonic electron as shown in Fig.~\ref{fig:HBDResponse:1}
is defined as a {\it{single cluster}}.
On the other hand,
a cluster created by merging blobs of $\rm{\check{C}}$erenkov light from
a track pair of photonic electrons as shown in Fig.~\ref{fig:HBDResponse:2}
is defined as a {\it{merging cluster}}.
However, a portion of the photonic electrons 
which have a large enough opening angle such that the two cluster do
not merge (typically $\simgt0.1$~rad)
creates two separated single clusters as shown
in Fig.~\ref{fig:HBDResponse:3}.
Therefore the single clusters are created by both of the nonphotonic
electron and the photonic electron with a large opening angle.

We also define another type of cluster created by scintillation light, 
which we call a {\it{scintillation cluster}}.
Scintillation hits which accidentally have large hit charges and
have neighboring hits can constitute clusters.
Photonic electrons from $\gamma$ conversions after the HBD GEM
pads do not create $\rm{\check{C}}$erenkov light in the HBD gas
volume.
Hence they basically do not have associated clusters in the HBD and they
are rejected by the HBD hit requirement in the eID-Cut.
However, a portion of these are accidentally associated with
scintillation clusters and satisfy the eID-Cut and so also survive in 
the reconstructed electron sample.

In Sec.~\ref{subsubsec:HBDClusterChargeFit}, we estimated
yields of these clusters from the distribution shape of the HBD cluster 
charge.
We also estimated the small component of {\it{single clusters}} generated
from photonic electrons which have the large opening angles
as described in Sec.~\ref{subsubsec:HBDRingChargeFit}.
Then we determined the nonphotonic electron yield.
Subtracting additional background electrons from $K_{e3}$ decays and
$e^{+}e^{-}$ decays of light vector mesons, we obtain the heavy
flavor electron yield as described
in Sec.~\ref{subsubsec:SingleE}.

\subsubsection{\label{subsubsec:HBDClusterChargeFit} Yield estimation of single clusters}


All clusters associated with the
reconstructed electrons can be classified into the above three types.
The yield of the electrons associated with the
single clusters must be evaluated to estimate the yield of the heavy flavor
electrons. 
The shapes of the $q_{\rm{clus}}$ distributions for the three cluster
types are quite different since merging  
clusters have basically double the charge of single clusters and
the charge of scintillation clusters is considerably smaller than the
charge of the single cluster.
Using the difference in the shapes, we estimate yields of these
clusters as follows.

The probability distributions of $q_{\rm{clus}}$ for single and
merging clusters were estimated by using low-mass unlike-sign electron
pairs reconstructed with only the eID-Cut,
which is dominated by photonic electron pairs.
We defined the unlike-sign electron pairs whose two electrons were associated
with two different HBD clusters as separated electron pairs and the pairs
whose two electrons were associated to the same HBD cluster as merging
electron pairs.
The probability distribution of $q_{\rm{clus}}$
for the single clusters were estimated by the
$q_{\rm{clus}}$ distribution of the separated electron pairs and
the probability distribution of $q_{\rm{clus}}$
for the merging clusters were estimated by the
$q_{\rm{clus}}$ distribution of the merging electron pairs.
The reconstruction of the electron pairs creates a small bias on the
shapes of the $q_{\rm{clus}}$ distributions.
Corrections for this bias are estimated by simulation and applied to
the distributions.
The probability distributions are denoted as
$f_{\rm{c}}^{\rm{s}}(q_{\rm{clus}})$ for the 
single clusters and $f_{\rm{c}}^{\rm{m}}(q_{\rm{clus}})$ for the merging
clusters.
The probability distribution of $q_{\rm{clus}}$
for the scintillation clusters is also
estimated by the distribution of the hadron tracks reconstructed by the
DC/PC tracking and the RICH veto and
denoted as $f_{\rm{c}}^{\rm{sci}}(q_{\rm{clus}})$.

The variables used in the {\bf{hbdcharge}} analysis are:
 
\begin{packed_descr}
\item[$f_{\rm{c}}^{\rm{s}}$] 
	  Probability distribution of $q_{\rm{clus}}$ 
	  for the single clusters 
\item[$f_{\rm{c}}^{\rm{m}}$]
	  Probability distribution of $q_{\rm{clus}}$ 
	  for the merging clusters 
\item[$f_{\rm{c}}^{\rm{sci}}$]
	  Probability distribution of $q_{\rm{clus}}$ 
	  for the scintillation clusters 
\item[$n_{\rm{s}}$]
	  Number of single clusters 
	  after applying eID-Cut. 
\item[$n_{\rm{m}}$]
	  Number of merging clusters 
	  after applying eID-Cut. 
\item[$n_{\rm{sci}}$]
	  Number of scintillation clusters 
	  after applying eID-Cut. 
\item[$\tilde{n}_{\rm{s}}$]
	  Number of single clusters 
	  after applying eID-Cut and npe-Cut.
\item[$\tilde{n}_{\rm{m}}$]
	  Number of merging clusters 
	  after applying eID-Cut and npe-Cut. 
\item[$\tilde{n}_{\rm{sci}}$]
	  Number of scintillation clusters 
	  after applying eID-Cut and npe-Cut. 
\end{packed_descr}

The $q_{\rm{clus}}$ distribution of the
reconstructed electrons found by applying eID-Cut is fitted with a
superposition of the three probability distributions
\begin{equation}
  \begin{array}{cc}
   n_{\rm{s}}\times f_{\rm{c}}^{\rm{s}}(q_{\rm{clus}})&+ \\
   n_{\rm{m}}\times f_{\rm{c}}^{\rm{m}}(q_{\rm{clus}})&+ \\
   n_{\rm{sci}}\times f_{\rm{c}}^{\rm{sci}}(q_{\rm{clus}}) ,&
  \end{array}
\end{equation}
where $n_{\rm{s}}$, $n_{\rm{m}}$ and $n_{\rm{sci}}$
are fitting parameters that represent the numbers of the
reconstructed electrons associating to
single clusters, merging
clusters and scintillation clusters
after applying eID-Cut respectively.
The fraction of nonphotonic electrons and photonic electrons are
different in different $p_{T}$ region of the reconstructed electron sample.
Therefore the fitting was performed for each $p_{T}$ region and
$n_{\rm{s}}(p_{T})$, $n_{\rm{m}}(p_{T})$ and
$n_{\rm{sci}}(p_{T})$ for each 
$p_{T}$ region were determined.
In the fitting, the distribution functions,
$f_{\rm{c}}^{\rm{s}}(q_{\rm{clus}})$,
$f_{\rm{c}}^{\rm{m}}(q_{\rm{clus}})$, and
$f_{\rm{c}}^{\rm{sci}}(q_{\rm{clus}})$, are assumed to be $p_{T}$
independent because the velocity of electrons in
$p_{T}$ region of interest
is close enough to the speed of light in vacuum such
that the yield of $\rm{\check{C}}$erenkov light from the electron is
nearly independent of $p_{T}$.
We also compared the shapes of the distributions in different
$p_{T}$ regions to confirm that the effect
from the track curvature is 
small enough to be ignored even at $p_{T}\sim0.5$~GeV/$c$.
On the other hand, $f_{\rm{c}}^{\rm{s}}(q_{\rm{clus}})$,
$f_{\rm{c}}^{\rm{m}}(q_{\rm{clus}})$ and
$f_{\rm{c}}^{\rm{sci}}(q_{\rm{clus}})$ 
for different HBD sectors vary slightly.
Considering this difference, the fitting is performed for each sector
individually.

\begin{figure}[tbh]
    \includegraphics[width=1.0\linewidth,clip]{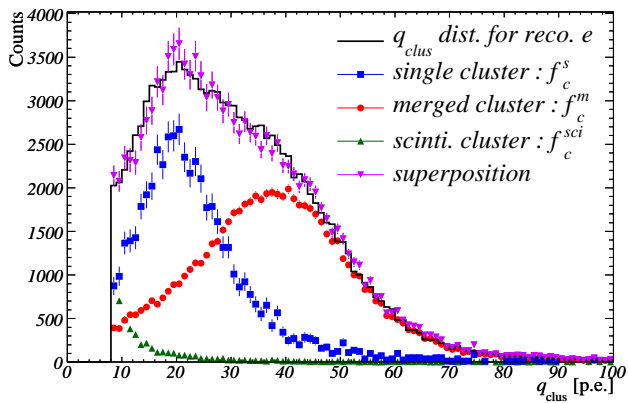}
    \caption{\label{fig:HBDClusterChargeFitting}(color online).
    Charge
    distribution of HBD clusters associated with reconstructed
    electrons with a transverse momentum ranging from $0.75$~GeV/$c$ to
    $1.00$~GeV/$c$ (solid black line),
    and the charge distribution for each
    component, i.e., single clusters
    ($n_{\rm{s}}f_{\rm{c}}^{\rm{s}}$,
    blue circles), merging clusters
    ($n_{\rm{m}}f_{\rm{c}}^{\rm{m}}$,
    red squares), and scintillation clusters
    ($n_{\rm{sci}}f_{\rm{c}}^{\rm{sci}}$,
    green triangles). The superposition
    of these components is also shown (purple inverted triangles).}
\end{figure}

The $q_{\rm{clus}}$ distribution for the reconstructed
electrons with transverse momentum $p_{T}$ ranging from $0.75$
GeV/$c$ to $1.00$ GeV/$c$ and the fitting result are shown in
Fig.~\ref{fig:HBDClusterChargeFitting} for one HBD sector.
The charge distribution of the reconstructed electrons is well
reproduced by the superposition of the three individual components.

\begin{figure}[tbh]
    \includegraphics[width=1.0\linewidth,clip]{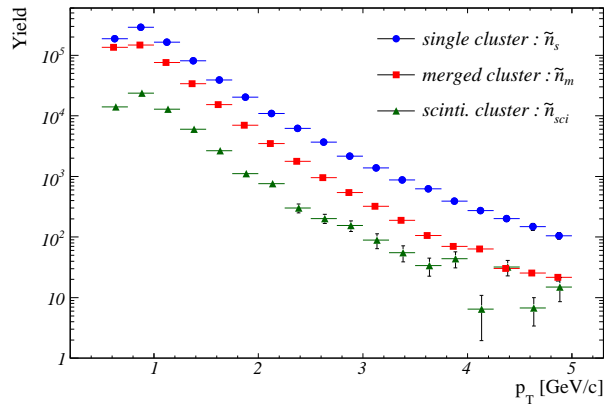}
    \caption{\label{fig:SepMergePtDistChargeCut}(color online).
    Yield spectra of HBD clusters after applying eID-Cut and npe-Cut
    estimated from the HBD cluster charge fitting.
    The plot shows the spectrum for the single clusters
    ($\tilde{n}_{\rm{s}}$, blue circle),
    the spectrum for the merging clusters
    ($\tilde{n}_{\rm{m}}$, red square) and
    the spectrum for the scintillation clusters
    ($\tilde{n}_{\rm{sci}}$, green triangle).
    The error bars represent fitting uncertainties.
    }
\end{figure}

The total number of reconstructed
electrons after applying both of eID-Cut
and npe-Cut for the three cluster types,
which are represented as $\tilde{n}_{\rm{s}}$,
$\tilde{n}_{\rm{m}}$ and $\tilde{n}_{\rm{sci}}$, are
calculated by applying the npe-Cut efficiencies of
$\int_{q_{{\rm min}}}^{q_{{\rm max}}}dqf^{\rm{s}}_{\rm{c}}(q)$,
$\int_{q_{{\rm min}}}^{q_{{\rm max}}}dqf^{\rm{m}}_{\rm{c}}(q)$ and
$\int_{q_{{\rm min}}}^{q_{{\rm max}}}dqf^{\rm{sci}}_{\rm{c}}(q)$
to the fit results, $n_{\rm{s}}$,
$n_{\rm{m}}$ and $n_{\rm{sci}}$, respectively.
In the integrals,
$q_{{\rm min}}$ and $q_{{\rm max}}$ represent the HBD charge
boundaries in the npe-Cut of $8$ p.e. and $28$ p.e. ($4$
p.e. and $17$ p.e. for the low-gain sector).
The variables, $\tilde{n}$, are also summarized above.
Figure~\ref{fig:SepMergePtDistChargeCut} shows the yield spectra from
the calculation as functions of $p_{T}$.

 \subsubsection{\label{subsubsec:HBDRingChargeFit} Yield estimation of
 separated photonic electrons}

The estimated $\tilde{n}_{\rm{s}}$ is the sum of
nonphotonic electrons and photonic electrons which
create the separated clusters in the HBD.
In the following description, we denote the photonic electrons which
create merging clusters as {\it{merging photonic electrons}} (MPE)
and those which
create separated single clusters as {\it{separated photonic
electrons}} (SPE).
In this section,
the number of SPE is estimated to obtain the yield of the nonphotonic
electrons.

\begin{figure}[tbh]
    \includegraphics[width=0.95\linewidth,clip]{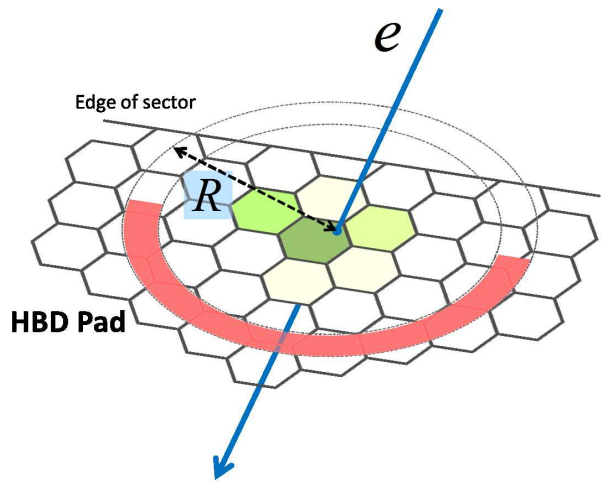}
    \caption{\label{fig:HBDRing}(color online).
    A half of an annular region around the reconstructed electron
    track on the HBD for the definition of {\bf{hbdringcharge}}.
    The inner and outer radii of the annular region
    are 7.0~cm and 8.0~cm respectively.
    The direction of the half region is determined as the opposite
    side to the edge of the HBD sector to avoid inefficiency around
    the edge.
    }
\end{figure}

\begin{figure}[tbh]
    \includegraphics[width=1.0\linewidth,clip]{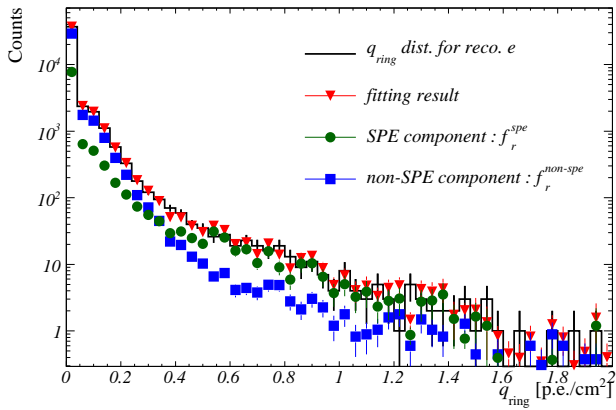}
    \caption{\label{fig:HBDRingChargeFitting}(color online). HBD charge
    distribution in the annular region for the reconstructed electrons
    with a transverse momentum ranging from 0.75~GeV/$c$ to 1.00~GeV/$c$
    (solid black line), and the fitting
    result of the charge
    distribution for electrons with correlated
    charges ($n_{\rm{spe}}f_{\rm{r}}^{\rm{spe}}$, green circle) and without
    correlated charges
    ($n_{\rm{non\mathchar`-spe}}f_{\rm{r}}^{\rm{non\mathchar`-spe}}$,
    blue square), and
    the superposition of the fitting results (red inverted triangle).
    }
\end{figure}

In the case where a reconstructed electron track is identified 
as an SPE, the partner electron generates an additional signal 
in the HBD, as illustrated in Fig.~\ref{fig:HBDResponse:3}. 
This property is utilized to estimate the number of SPE. For 
this estimation, we defined a new value, $q_{\rm{ring}}$, as

\begin{description}
 \item[hbdringcharge: $q_{\rm{ring}}$] 
The total charge in the HBD
	      pads centered on a half of an annular region with an inner
	      radius of $7.0$~cm and an outer radius of $8.0$~cm around the track
	      projection of HBD as shown in Fig.~\ref{fig:HBDRing}.
	      To avoid inefficient regions around the edges of the HBD
	      sectors, we use one half of an annular region oriented
	      away from the nearest sector edge (see Fig.~\ref{fig:HBDRing}).
	      The $q_{\rm{ring}}$ value is 
	      normalized by the area of the half of
	      the annular region in the definition.
\end{description}

The choice of 7.0~cm to 8.0~cm is determined by three factors: 
(1) the distribution of distance between separated clusters of 
SPE has a maximum around 7.0~cm, (2) few HBD clusters have 
radii larger than 7.0~cm, and (3) larger area includes more 
scintillation background and decreases the signal to background 
ratio. Whereas the $q_{\rm{ring}}$ distributions for the 
nonphotonic electrons and MPE comprise signals only from 
scintillation light, the distributions for SPE include the 
correlated signals around the tracks in addition to 
scintillation light.

The variables used in the {\bf{hbdringcharge}} analysis are: 
  
\begin{packed_descr}
\item[$f_{\rm{r}}^{\rm{spe}}$]
	  Probability distribution of $q_{\rm{ring}}$ for SPE 
\item[$f_{\rm{r}}^{\rm{non\mathchar`-spe}}$]
	  Probability distribution of $q_{\rm{ring}}$ 
	  for nonphotonic electrons and MPE 
\item[$n_{\rm{spe}}$]
	  Number of SPE 
	  after applying the eID-Cut and npe-Cut. 
\item[$n_{\rm{non\mathchar`-spe}}$]
	  Number of electrons other than SPE 
	  after applying the eID-Cut and npe-Cut. 
\end{packed_descr}

The $f_{\rm{r}}^{\rm{spe}}(q_{\rm{ring}})$ for SPE and the 
$f_{\rm{r}}^{\rm{non\mathchar`-spe}}(q_{\rm{ring}})$ for 
nonphotonic electrons and MPE can be estimated by hadron tracks 
and electron tracks with large $q_{\rm{clus}}$ values, which 
consist almost entirely of MPE.  Because hadrons and MPE 
clusters do not create any correlated signals around their 
tracks, the $q_{\rm{ring}}$ distributions of the tracks are 
created by only the scintillation light.

The $f_{\rm{r}}^{\rm{spe}}(q_{\rm{ring}})$ was estimated by 
using simulations. The dominant photonic electrons come from 
the Dalitz decays of $\pi^{0}$ and $\eta$ and $\gamma$ from 
their decays which convert in materials. We simulated the 
detector responses for the Dalitz decay and the $\gamma$ 
conversion events of the neutral mesons by a {\sc geant}3 
simulation \cite{GEANT3:1994} configured for the PHENIX 
detector system. The $\pi^{0}$ and $\eta$ spectra were 
parametrized in the simulation by $m_{T}$-scaled Tsallis 
distributions \cite{MTScaling:PRD2011}, together with their 
known branching ratios to Dalitz decays and $\gamma$ decays. In 
order to include contributions from scintillation light, 
$f_{\rm{r}}^{\rm{non\mathchar`-spe}}(q_{\rm{ring}})$, which is 
identical to the $q_{\rm{ring}}$ distribution from only the 
scintillation light, was convoluted to the result to obtain 
$f_{\rm{r}}^{\rm{spe}}(q_{\rm{ring}})$.

The $q_{\rm{ring}}$ distribution for the reconstructed
electrons selected by applying eID-Cut and npe-Cut
was fitted with the superposition 
of the $q_{\rm{ring}}$ distributions,
$f_{\rm{r}}^{\rm{spe}}(q_{\rm{ring}})$
and $f_{\rm{r}}^{\rm{non\mathchar`-spe}}(q_{\rm{ring}})$, as
\begin{eqnarray}
 n_{\rm{spe}}\times f_{\rm{r}}^{\rm{spe}}(q_{\rm{ring}})+
 n_{\rm{non\mathchar`-spe}}\times f_{\rm{r}}^{\rm{non\mathchar`-spe}}(q_{\rm{ring}}) ,
\end{eqnarray}
where $n_{\rm{spe}}$ and $n_{\rm{non\mathchar`-spe}}$ are fitting
parameters and represent the numbers of SPE and other electrons
in the $q_{\rm{ring}}$ distribution, respectively,
as summarized above.
Similar to the $q_{\rm{clus}}$ distribution,
the fitting for the $q_{\rm{ring}}$ distribution was also performed
for each electron $p_{T}$ region and each HBD sector.
Figure~\ref{fig:HBDRingChargeFitting} shows a fitting result in one HBD
sector in the electron $p_{T}$ region from $0.75$ GeV/$c$ to $1.00$
GeV/$c$.

\subsubsection{\label{subsubsec:SingleE}Yield estimation of heavy flavor electrons}

Using the above fitting results of $\tilde{n}_{\rm{s}}$
and $n_{\rm{spe}}$,
the yield of nonphotonic electrons, $N^{\rm{npe}}$ was
estimated with the formula
\begin{eqnarray}
 \label{eq:NonPhotonicEFormula}
  N^{\rm{npe}}(p_{T}) & = &
  \tilde{n}_{\rm{s}}(p_{T}) - n_{\rm{spe}}(p_{T}). 
\end{eqnarray}
The remaining background for the heavy flavor electrons 
in the nonphotonic electron sample comes from $K_{e3}$
decays and $e^{+}e^{-}$ decays of light vector
mesons, namely $\rho$, $\omega$, and $\phi$.
Electrons from the Drell-Yan process also contribute to the
background, however the contribution is known to be less
than 0.5\% of total heavy flavor electrons in this $p_{T}$ range and
can be ignored. 
We determined the yield of the
heavy flavor
electrons from
$N^{\rm{npe}}$ by subtracting the components of the $K_{e3}$
electrons, which are estimated by simulation using a measured $K$ cross
section \cite{MTScaling:PRD2011}, and the electrons from light vector
mesons, which are already estimated in previously published
result \cite{SingleE:PRL2006}, as 
\begin{eqnarray}
 \label{eq:SingleEFormula}
 N^{{\rm{HFe}}}(p_{T}) = N^{\rm{npe}}(p_{T}) - N^{Ke3}(p_{T}) -
 N^{\rm{LVM}}(p_{T}),  
\end{eqnarray}
where $N^{Ke3}(p_{T})$ and $N^{\rm{LVM}}(p_{T})$ represent the
electrons from the $K_{e3}$ decays and the light vector meson decays
respectively.

\subsection{\label{subsec:SystematicErrorInMSE}Systematic Uncertainty}

\begin{table}
 \caption{\label{tab:YieldSysError} Relative systematic uncertainties
 given on the heavy flavor electron yield.}
 \begin{ruledtabular}\begin{tabular}{c c l}
  source & uncertainty & \ \ \ $p_{T}$ range (GeV/$c$) \\
\hline
 {\bf{hbdringcharge}} fitting & $16\%$ & ( $0.50<p_{T}<0.75$ ) \\
                              & $ 6\% \sim 4\%$ & ( $0.75<p_{T}<1.75$ ) \\
                              & $ 2\%$ & ( $1.75<p_{T}$ ) \\
 {\bf{hbdcharge}} fitting & $2\%$ & ( $0.50<p_{T}<0.75$ ) \\
                          & $< 1\%$ & ( $0.75<p_{T}$ ) \\
 $K_{e3}$ & $4\%$ & ( $0.50<p_{T}<0.75$ ) \\
          & $< 1\%$ & ( $0.75<p_{T}$ ) \\
 hadron misID & $4\%$ & ( $0.50<p_{T}<0.75$ ) \\
               & $< 1\%$ & ( $0.75<p_{T}$ ) \\
\end{tabular}\end{ruledtabular}
\end{table}

The systematic uncertainties for the heavy flavor electron yield come
from the fits for
the $q_{\rm{clus}}$ distribution and
the $q_{\rm{ring}}$ distribution,
and from estimations of $K_{e3}$
contribution and misidentified hadrons.

The most significant source in these contributions is the fitting
uncertainty for the $q_{\rm{ring}}$ distribution.
We varied the radius of the annular region to
an inner radius of $6.0$ cm and an
outer radius of $7.0$ cm and also to $8.0$ cm and $9.0$ cm from
the default radii of $7.0$ cm and $8.0$ cm.
The uncertainty from the fitting was set to the amount of 
variation in $n_{\rm{spe}}$ after these changes.
The estimated uncertainties decrease from
about 16\% of the heavy flavor electron yield in
the momentum range of $0.50<p_{T}<1.00$~GeV/$c$ to about 2\% above
1.75~GeV/$c$.

The fitting uncertainty for the $q_{\rm{clus}}$ distribution
comes from the estimation of the bias in the charge distribution shape
due to the electron pair reconstruction.
The systematic uncertainty from this effect is estimated to be
less than 2\% by simulation.

In the low momentum region, $0.50<p_{T}<1.00$~GeV/$c$,
uncertainties from the $K_{e3}$ contribution
and the hadron misreconstruction are significant.
The uncertainty from the $K_{e3}$ contribution comes almost entirely 
from the uncertainty on the
$K$ cross section used in the $K_{e3}$ simulation.
This uncertainty amounts to about 4\% of the total
heavy flavor electron yield in the low
momentum region and decreases to less than 1\% for $p_{T}>0.75$~GeV/$c$.
We also estimated the upper limits of the hadron contamination due to
misreconstructions employing a hadron-enhanced event set.
As a result, we determined the upper limits as 4\% of the total
heavy flavor electron yield in the low
momentum region which decreases to less than 1\% over 1.5~GeV/$c$.
The upper limits are assigned as the systematic uncertainties from hadron
misreconstructions.
Table~\ref{tab:YieldSysError} summarizes the systematic uncertainties
on the heavy flavor electron yield.

\subsection{\label{subsec:SEResult}Results of Heavy Flavor Electron Yield and
Signal Purity}

\begin{figure}[tbh]
    \includegraphics[width=1.0\linewidth,clip]{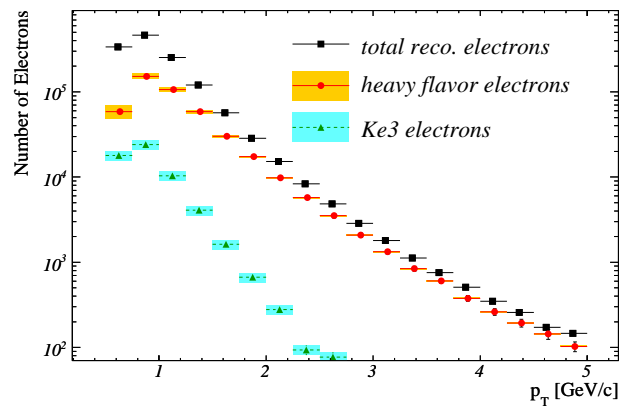}
    \caption{\label{fig:SingleE}(color online). Heavy flavor electron yield
    spectrum. The black square points represent the total number of the
    reconstructed electrons after applying the eID-Cut and npe-Cut.
    The red circle points represent the
    estimated yield of the heavy flavor electrons. The yellow bands
    represent the systematic uncertainties for the heavy flavor
    electron yield. 
    The green triangle points with dashed lines
    represent the estimated $K_{e3}$
    contribution with systematic uncertainties shown by light-blue bands.}
\end{figure}

From Eq.~\ref{eq:NonPhotonicEFormula} and
Eq.~\ref{eq:SingleEFormula} and the discussion in
Sec.~\ref{subsec:SystematicErrorInMSE}, the 
heavy flavor electron yield spectrum with the systematic uncertainties was
determined. 
The spectrum is shown in Fig.~\ref{fig:SingleE}.
We also show the yield of inclusive reconstructed electrons after
applying the eID-Cut and npe-Cut and the estimated $K_{e3}$
contribution.
The electrons from $e^{+}e^{-}$ decays of the light vector mesons are
not shown in Fig.~\ref{fig:SingleE}, but they are less than 5\%
of the heavy flavor electron yield in this $p_{T}$ range. 

\begin{figure}[tbh]
    \includegraphics[width=1.0\linewidth,clip]{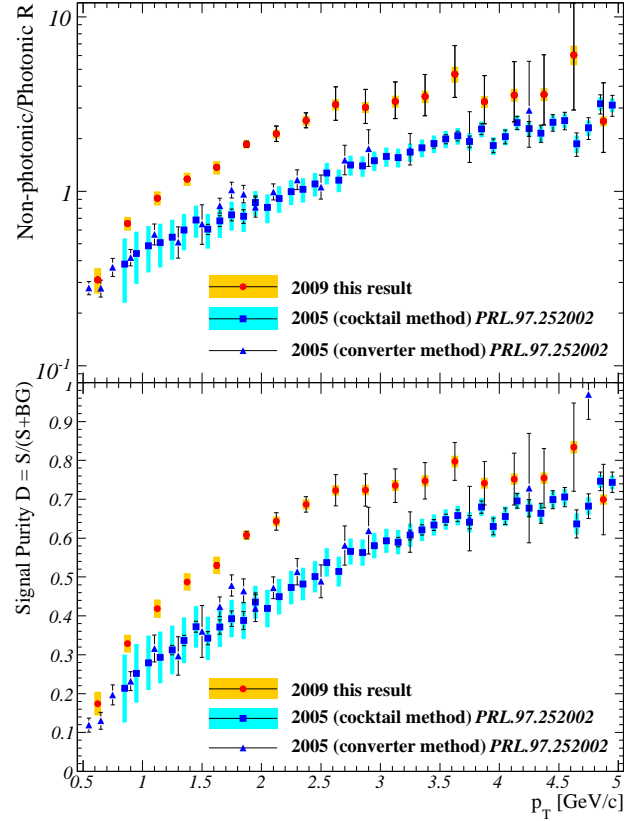}
    \caption{\label{fig:SignalFraction}(color online). (top) Ratio
    between the yields of the nonphotonic electrons and the photonic
    electrons in the reconstructed tracks. The red circles and the
    blue squares and triangles
    represent this analysis result and the previous result \cite{SingleE:PRL2006},
    respectively. The error bars and bands
    represent the statistic and the systematic uncertainties.
    (bottom) Signal purity which is a ratio of the yield of the
    heavy flavor electrons to the total reconstructed electrons.}
\end{figure}

The ratio of the nonphotonic electron yield 
to the photonic electron
yield in this measurement,
\begin{eqnarray}
 \label{eq:NPEPERatio}
 R(p_{T})\equiv \frac{N^{\rm{npe}}(p_{T})}{N^{\rm{reco}}_{e}(p_{T})-N^{\rm{npe}}(p_{T})}
\end{eqnarray}
where $N^{\rm{reco}}_{e}$ denotes the total number of  
reconstructed electrons after applying the eID-Cut and npe-Cut,
is shown in the top panel of Fig.~\ref{fig:SignalFraction}.
In Eq.~\ref{eq:NPEPERatio}, we assumed the fraction of misidentified
hadrons in the reconstructed electrons after the cuts is negligible as
shown in Fig.~\ref{fig:EPRatio},
and so the number of photonic electrons can be represented as
$N^{\rm{reco}}_{e}(p_{T})-N^{\rm{npe}}(p_{T})$.
The same ratio from a previous
measurement \cite{SingleE:PRL2006} is also shown in the figure.
The previous measurement employed two other methods for the background
estimation, namely a cocktail method and a converter method.
In the cocktail method,
a sum of electron spectra from various background sources was
calculated using a Monte Carlo hadron decay generated.
This sum was subtracted from the inclusive electron sample to isolate
the heavy flavor contribution.
With the converter method, a photon converter around the beam pipe was
introduced to increase the photon conversion probability by a
well-defined amount, and thus allow determination of the photonic
background.
The nonphotonic to photonic electron ratio 
is improved by a factor of about 2 or more in
$p_{T}>1.0$~GeV/$c$ compared with the previously measured result
due to the rejection of photonic electrons by the HBD.

The signal purity is defined as the ratio of the yield of the
heavy flavor electrons to the reconstructed electrons after applying the 
eID-Cut and npe-Cut,
\begin{eqnarray}
 \label{eq:SignalPurity}
 D(p_{T})\equiv \frac{N^{\rm{HFe}}(p_{T})}{N^{\rm{reco}}_{e}(p_{T})}.
\end{eqnarray}
The result is shown as the bottom plot in Fig.~\ref{fig:SignalFraction}.
We also show the result of the signal purity in the previous
measurement.
Comparing with the previously measured result,
the signal purity is improved by
a factor of about $1.5$ in a $p_{T}$ range from $0.75$~GeV/$c$ to
$2.00$~GeV/$c$. 

 \section{\label{sec:CrossSection}Heavy Flavor Electron Cross Section}

\begin{table}
 \caption{\label{tab:XSectSysError}
 Relative systematic uncertainties on
 the cross section due to uncertainties in the total sampled luminosity,
 trigger efficiencies, and detector acceptance.
 These systematic uncertainties
 are globally correlated in all $p_{T}$ regions
 ($p_{T}>1.25$~GeV/$c$ for
 the uncertainties on $\epsilon_{{\rm trig}}^{e\rm{|MB}}$).}
\begin{ruledtabular}\begin{tabular}{c c l}
 source & uncertainty &  \\
\hline 
 MB trig. cross sect. & $9.6\%$ &  \\
 acceptance $A$ & $8\%$   &  \\
 reco. efficiency $\epsilon_{{\rm rec}}$ & $6\%$   &  \\
 MB trig. efficiency $\epsilon_{{\rm trig}}^{{\rm MB}}$ & $2.5\%$ &  \\
 $e$ trig. efficiency $\epsilon_{{\rm trig}}^{e\rm{|MB}}$ & $3.6\%$ & in $p_{T}>1.25$ GeV/$c$ \\
\end{tabular}\end{ruledtabular}
\end{table}
	
The invariant cross section is calculated from
\begin{equation}
 E\frac{d^{3}\sigma}{dp^{3}} = \frac{1}{2\pi
p_{T}}\frac{1}{L}\frac{1}{A\epsilon_{{\rm rec}}\epsilon_{{\rm trig}}}\frac{N(\Delta
p_{T},\Delta y)}{\Delta
p_{T}\Delta y} ,
\end{equation}
where $L$ denotes the integrated luminosity, $A$ the acceptance,
$\epsilon_{{\rm rec}}$ the reconstruction efficiency,
$\epsilon_{{\rm trig}}$ the trigger efficiency, and $N$ the estimated
number of heavy flavor electrons.

The luminosity, $L$, was calculated from the number of MB events divided by the
cross section for the MB trigger.
For the latter, a value of $23.0$~mb with a systematic uncertainty of
$9.6$\% was estimated from van-der-Merr scan
results \cite{Pi0pp:PRL2003} corrected for the relative changes in the BBC
performance. 
The combination of the acceptance and the reconstruction efficiency,
$A\epsilon_{{\rm rec}}(p_{T})$, was estimated by a {\sc geant}3 simulation.
We found that $A\epsilon_{{\rm rec}}(p_{T})$ has a value of
$4.7\%\times(1\pm 8\times10^{-2}(\rm{acc.})\pm
6\times10^{-2}(\rm{reco.}))$, with a slight $p_{T}$ dependence.

\begin{figure}[tbh]
    \includegraphics[width=1.0\linewidth,clip]{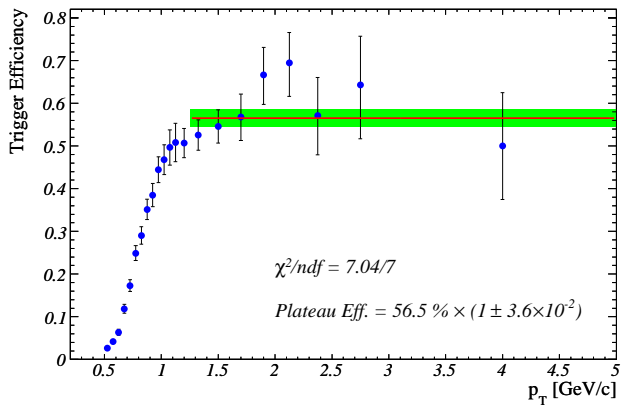}
    \caption{\label{fig:TriggerEfficiency}(color online). Efficiency of the
    electron trigger for reconstructed electrons under the
    condition that the MB trigger was issued. The red line represents
    the fitting result with the constant function and the green band
    represents the fitting uncertainty.}
\end{figure}

The efficiency of the MB trigger for the hard scattering processes,
including heavy flavor electron production, is
$\epsilon_{{\rm trig}}^{{\rm MB}} = 79.5\% \times (1 \pm 2.5\times10^{-2})$.
The efficiency of the electron trigger for the electrons under the
condition of the MB trigger firing,
$\epsilon_{{\rm trig}}^{e\rm{|MB}}(p_{T}) \equiv \epsilon_{{\rm trig}}(p_{T})/\epsilon_{{\rm trig}}^{{\rm MB}}$,
can be calculated by the ratio of the
number of the reconstructed electrons in the MB triggered sample in
coincidence with the electron trigger to the number of the reconstructed
electrons without the coincidence.
The efficiency $\epsilon_{{\rm trig}}^{e\rm{|MB}}$
is shown in Fig.~\ref{fig:TriggerEfficiency} as a function of
$p_{T}$.
Whereas we used the calculated efficiency values for the momentum
region of $p_{T}<1.25$ GeV/$c$,
we assumed a saturated efficiency for $p_{T}>1.25$ GeV/$c$ and estimated
the value with a fitting as shown in Fig.~\ref{fig:TriggerEfficiency}.
The fitting result is $\epsilon_{\rm{plateau}}=56.5\%\times (1\pm
3.6\times10^{-2})$. 
The total trigger efficiency $\epsilon_{{\rm trig}}(p_{T})$ can be
calculated with the above two efficiencies as
$\epsilon_{{\rm trig}}(p_{T}) = \epsilon_{{\rm trig}}^{{\rm MB}}\times
\epsilon_{{\rm trig}}^{e\rm{|MB}}(p_{T})$.
Table~\ref{tab:XSectSysError} summarizes the systematic uncertainties on
the cross section due to uncertainties in the total sampled luminosity,
trigger efficiencies, and detector acceptance.
All systematic uncertainties listed in Table~\ref{tab:XSectSysError}
are globally correlated over whole $p_{T}$ region ($p_{T}>1.25$~GeV/$c$ for
the uncertainties on $\epsilon_{{\rm trig}}^{e\rm{|MB}}$).

\begin{figure}[tbh]
    \includegraphics[width=1.0\linewidth,clip]{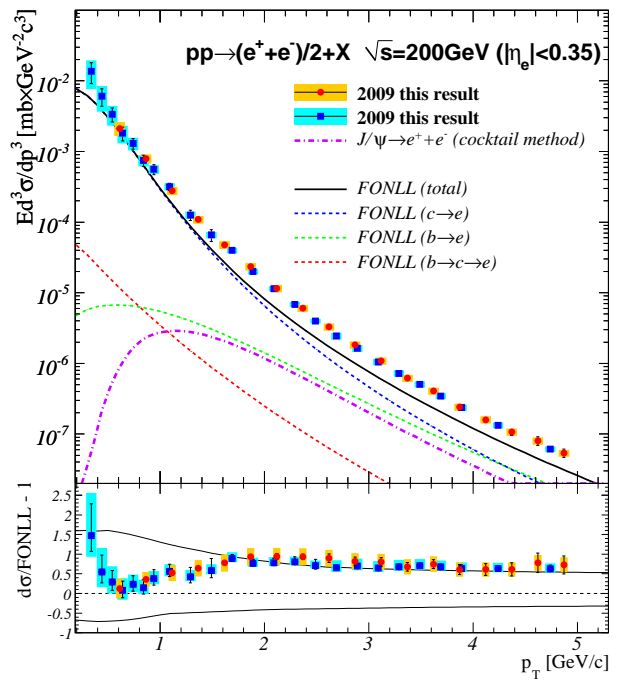}
    \caption{\label{fig:CrossSection}(color online). (top) Invariant
    differential cross sections of electrons from heavy-flavor
    decays.  The red circles are 
    this analysis of 2009 data and the blue squares are the previous 
    2005 data~\protect\cite{SingleE:PRL2006} for 
    the nonphotonic electron cross sections.
    The error bars and bands represent the statistical
    and systematic uncertainties.
    The scaling uncertainty from the Vernier scan is not included in
    the error bands because the same uncertainty must be
    considered for both the results of 2009 and 2005.
    The purple dashed dotted line is electron cross section from
    $J/\psi\rightarrow e^{+}+e^{-}$  decays estimated from the cocktail
    method~\protect\cite{HQPhenix:PRC2011}. 
    The solid and dashed curves are the FONLL
    calculations. (bottom) Difference of the ratio of
    the data and the FONLL
    calculation from 1.
    The upper and lower curve shows the theoretical
    upper and lower limit of the FONLL calculation.}
\end{figure}

The measured cross section of heavy flavor electrons is shown in
Fig.~\ref{fig:CrossSection} and tabulated in 
Table~\ref{tab:CrossSection}.
A correction for bin width \cite{BinningCorr:NIM1995} is applied to the
$p_{T}$ value of each point.
The figure also shows the previously published
result \cite{SingleE:PRL2006}.
The new result agrees well with the previous result within the
uncertainties.
Note that in this paper we employed a new analysis method with the HBD whereas the
previous measurement employed different methods,
the cocktail method and the converter method.
The consistency between these measurements proves that additional
photonic backgrounds generated in the HBD material are removed, and that
this new analysis method with the HBD is robust.

The electron cross section from $J/\psi\rightarrow e^{+}+e^{-}$ decays
estimated by the cocktail method \cite{HQPhenix:PRC2011}
and a fixed order next-to-leading log (FONLL) pQCD calculation of the
heavy flavor contributions to the electron
spectrum \cite{FONLOXSect:PRL2005} are also shown in Fig.~\ref{fig:CrossSection}.
The $J/\psi$ contribution to the heavy flavor electrons are less than
2\% in $p_{T}<1.25$~GeV/$c$ and increase to $\sim$20\% until
$p_{T}=5.0$~GeV/$c$. 
The FONLL pQCD calculation shows that
the heavy flavor electrons in the low momentum region are dominated by
charm quark decays, and the contribution from bottom quarks in
$p_{T}<1.25$~GeV/$c$ is less than 5\%. 
 
\begin{table}[tbh]
\caption{\label{tab:CrossSection}
Data table for the cross section result corresponding 
to Fig.~\ref{fig:CrossSection}.}
\begin{ruledtabular}\begin{tabular}{c c c c}
$p_{T}$ & $E\frac{d^{3}\sigma}{dp^{3}}$ & stat. error 
& syst. error \\
 {[GeV/$c$]} & \multicolumn{3}{c}{[mb$\times$GeV$^{-2}$c$^{3}$]} \\ 
\hline
0.612 & 2.12$\times10^{-3}$ & 0.04$\times10^{-3}$ & 0.47$\times10^{-3}$ \\
0.864 & 7.93$\times10^{-4}$ & 0.09$\times10^{-4}$ & 1.11$\times10^{-4}$ \\
1.115 & 2.78$\times10^{-4}$ & 0.03$\times10^{-4}$ & 0.37$\times10^{-4}$ \\
1.366 & 1.09$\times10^{-4}$ & 0.02$\times10^{-4}$ & 0.13$\times10^{-4}$ \\
1.617 & 4.77$\times10^{-5}$ & 0.08$\times10^{-5}$ & 0.58$\times10^{-5}$ \\
1.867 & 2.34$\times10^{-5}$ & 0.05$\times10^{-5}$ & 0.27$\times10^{-5}$ \\
2.118 & 1.15$\times10^{-5}$ & 0.04$\times10^{-5}$ & 0.13$\times10^{-5}$ \\
2.369 & 6.05$\times10^{-6}$ & 0.20$\times10^{-6}$ & 0.68$\times10^{-6}$ \\
2.619 & 3.28$\times10^{-6}$ & 0.19$\times10^{-6}$ & 0.37$\times10^{-6}$ \\
2.869 & 1.82$\times10^{-6}$ & 0.11$\times10^{-6}$ & 0.20$\times10^{-6}$ \\
3.120 & 1.08$\times10^{-6}$ & 0.07$\times10^{-6}$ & 0.12$\times10^{-6}$ \\
3.370 & 6.20$\times10^{-7}$ & 0.41$\times10^{-7}$ & 0.69$\times10^{-7}$ \\
3.620 & 4.07$\times10^{-7}$ & 0.26$\times10^{-7}$ & 0.45$\times10^{-7}$ \\
3.870 & 2.42$\times10^{-7}$ & 0.19$\times10^{-7}$ & 0.27$\times10^{-7}$ \\
4.121 & 1.59$\times10^{-7}$ & 0.15$\times10^{-7}$ & 0.18$\times10^{-7}$ \\
4.371 & 1.07$\times10^{-7}$ & 0.11$\times10^{-7}$ & 0.12$\times10^{-7}$ \\
4.621 & 8.02$\times10^{-8}$ & 1.11$\times10^{-8}$ & 0.89$\times10^{-8}$ \\
4.871 & 5.38$\times10^{-8}$ & 0.71$\times10^{-8}$ & 0.60$\times10^{-8}$ \\
\end{tabular}\end{ruledtabular}
\end{table}

 \section{\label{sec:HelicityAsymmetry}Heavy Flavor Electron Spin Asymmetry}

\begin{table}
 \caption{\label{tab:ALLSysError} Systematic uncertainties by type.
 The scaling uncertainty denotes an uncertainty on 
 scaling of the raw asymmetry $A_{LL}^{\rm{S+BG}}$ and the offset
 uncertainty denotes
 an uncertainty on the absolute value of the asymmetry.
 The ``global'' in this table means the uncertainties are globally
 correlated in all $p_{T}$ regions.
 The scaling uncertainty is represented as the ratio of the uncertainty
 to the signal ($\Delta S/S$) given in percent
 and the offset uncertainty is represented as the
 absolute value of the uncertainty.}
\begin{ruledtabular}\begin{tabular}{c c c}
 source & uncertainty & type \\
\hline 
 signal purity $D$ & $\sim6\%$ &  scaling \\
 polarization ( $\frac{\Delta(P_{B}P_{Y})}{P_{B}P_{Y}}$ ) & $8.8\%$ &
	   global scaling \\
 relative luminosity $r$ & \ $0.14\times 10^{-2}$ \  &  global offset \\
 background asymmetry $A_{LL}^{BG}$ & \ $0.2\times
     10^{-2}\times\frac{1-D}{D}$\  &  offset \\ 
\end{tabular}\end{ruledtabular}
\end{table}

Since parity is conserved in QCD processes, thereby disallowing finite
single spin asymmetries, using Eq.~\ref{eq:PolCrossSection} we express
the expected electron yields for each beam-helicity combination as
\begin{equation}
\label{eq:Yield}
\begin{array}{l}
 N_{++}^{\rm{exp}}(N_{0},A_{LL}) = N_{0}(1 + |P_{\rm{B}}P_{\rm{Y}}|A_{LL}) \\
 N_{--}^{\rm{exp}}(N_{0},A_{LL}) = N_{0}(1 + |P_{\rm{B}}P_{\rm{Y}}|A_{LL})/r_{--} \\
 N_{+-}^{\rm{exp}}(N_{0},A_{LL}) = N_{0}(1 - |P_{\rm{B}}P_{\rm{Y}}|A_{LL})/r_{+-} \\
 N_{-+}^{\rm{exp}}(N_{0},A_{LL}) = N_{0}(1 - |P_{\rm{B}}P_{\rm{Y}}|A_{LL})/r_{-+},
\end{array}
\end{equation}
where
$N_{\pm\pm}^{\rm{exp}}(N_{0},A_{LL})$
denote the expected yields for collisions between
the blue beam-helicity ($\pm$) and
the yellow beam-helicity ($\pm$) and $N_{0}$
is the expected yield in collisions of unpolarized beams under the same
integrated luminosity as the $++$ beam-helicity combination.
$N_{\pm\pm}^{\rm{exp}}(N_{0},A_{LL})$ are used for fitting functions
to estimate $A_{LL}$ as described below.
$P_{\rm{B}}$ and $P_{\rm{Y}}$ represent the polarizations of the beams.
The beam polarizations are measured with a carbon target
polarimeter \cite{pC:arXiv2004}, 
normalized by the absolute polarization measured with a separate
polarized atomic hydrogen jet polarimeter \cite{pH:PLB2006,pH:arXiv2007}
at another collision point in RHIC ring.
The measured polarizations are about
$P=57\%$ with a relative uncertainty of $\Delta
P/P=4.7\times10^{-2}$ in the measurement.  
The relative luminosities are defined as the ratio of the
luminosities in the beam-helicity combinations,
\begin{equation}
 \begin{array}{l}
r_{--} \equiv \frac{L_{++}}{L_{--}} \\
r_{+-} \equiv \frac{L_{++}}{L_{+-}} \\
r_{-+} \equiv \frac{L_{++}}{L_{-+}} ,
 \end{array}
\end{equation}
where $L_{\pm\pm}$ represent the integrated luminosities in the
beam-helicity combinations shown by the subscript.
The relative luminosities are determined by the ratios of MB trigger
counts in the four beam-helicity combinations.

The double-spin asymmetry for inclusive electrons after applying eID-Cut
and npe-Cut, which include not only
the heavy flavor electrons (S) but also the background electrons (BG),
is determined by simultaneously fitting the yields of electrons in each
of the four beam-helicity combinations with the expected values
$N_{\pm\pm}^{\rm{exp}}(N_{0},A_{LL})$ from Eq.~\ref{eq:Yield}, where
$A_{LL}$ and $N_{0}$ are free parameters.
To perform the fit, a log likelihood method assuming Poisson
distributions with expected values of $N_{\pm\pm}^{\rm{exp}}(N_{0},A_{LL})$
was employed.
The fit was performed
for electron yields in each fill
to obtain the fill-by-fill double-spin asymmetry.
We confirmed that all asymmetries in different fills are consistent with
each other within their statistical uncertainties and, therefore,
the patterns of the crossing helicities in the fills do not affect the
asymmetry measurement. 
The final double-spin asymmetry for inclusive electrons,
$A_{LL}^{\rm{S+BG}}(p_{T})$, was 
calculated as the weighted mean of the fill-by-fill asymmetries.

The double-spin asymmetry in the heavy flavor electron production,
$A_{LL}^{\rm{HFe}}$, was determined from 
\begin{equation}
\label{eq:ALL}
 A_{LL}^{\rm{HFe}}(p_{T})=\frac{1}{D(p_{T})}A_{LL}^{\rm{S+BG}}(p_{T})-\frac{1-D(p_{T})}{D(p_{T})}A_{LL}^{\rm{BG}}(p_{T})  
\end{equation}
where $A_{LL}^{\rm{BG}}$ represents the spin
asymmetries for the background electron production,
and $D$ represents the signal purity defined in Eq.~\ref{eq:SignalPurity}
and shown in Fig.~\ref{fig:SignalFraction}.
As previously discussed, most of the background electrons come from
Dalitz decays of the $\pi^{0}$ and $\eta$, or from conversions of
photons from decays of those hadrons.
The fractional contribution on the partonic level, and therefore the
production mechanism for the $\pi^{0}$ and $\eta$ is expected to be very
similar up to
$\sim10$~GeV/$c$ \cite{PHENIXPi0ALL:PRD2007,PHENIXEtaALL:PRD2011}.
We assume identical spectra for
double-spin asymmetries of $\pi^{0}$ production and $\eta$ production,
and estimated $A_{LL}^{\rm{BG}}$ from only the $\pi^{0}$ double-spin
asymmetry using data from this PHENIX measurement.
The resulting $A_{LL}^{\rm{BG}}$ is
$-0.1\times10^{-2}<A_{LL}^{\rm{BG}}<0.1\times10^{-2}$ 
in $0.5<p_{T}<2.5$ GeV/$c$ and
$0.1\times10^{-2}<A_{LL}^{\rm{BG}}<0.2\times10^{-2}$ in 
$2.5<p_{T}<3.0$ GeV/$c$, with uncertainties less than
$0.2\times10^{-2}$. 

Systematic uncertainties on $A_{LL}^{\rm{HFe}}$ are separated into
scaling uncertainties and offset uncertainties.
The scaling uncertainties come from uncertainty in the beam
polarizations, $P_{B}$ and $P_{Y}$, and the signal purity, $D$.
The uncertainty from the beam polarization is estimated as
$\Delta(P_{B}P_{Y})/P_{B}P_{Y}=8.8$\% which is globally correlated over
the whole $p_{T}$ range.
The offset uncertainties come from uncertainties in the relative
luminosity, $r$, and the background asymmetry, $A_{LL}^{\rm{BG}}$.
The uncertainty from relative luminosity which is also globally correlated over
is determined as
$\Delta r=1.4\times10^{-3}$ from comparison of the measured relative
luminosities with the MB trigger and the
ZDC trigger.
The systematic uncertainties are summarized in
Table~\ref{tab:ALLSysError}.

A transverse double-spin asymmetry $A_{TT}$, which is defined by the same
formula as Eq.~\ref{eq:ALLDef} for the transverse polarizations, can
contribute to $A_{LL}$ through the residual transverse components of the
beam polarizations.
The product of the transverse components of the beam polarization is
measured to be $\sim10^{-2}$ in this experiment.
For $\pi^{0}$ production, the $A_{TT}$ is expected to be
$\sim10^{-4}$ based on an NLO QCD
calculation \cite{Pi0ATT:PRD2005}.
If we assume the transverse asymmetries of $\pi^{0}$ and heavy flavor
electrons are comparable, we arrive at the value of $A_{LL}\sim10^{-6}$.
This value is negligible compared with the precision of the
$A_{LL}^{S+BG}$ measurement of $\sim10^{-3}$.

\begin{figure}[tbh]
\includegraphics[width=1.0\linewidth,clip]{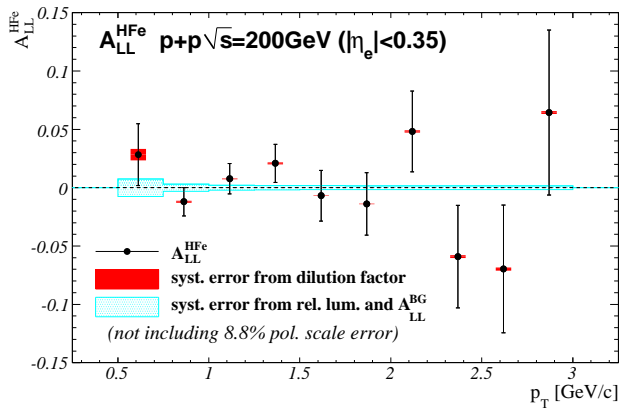}
\caption{\label{fig:SingleEALL}(color online). Double-spin asymmetry
of the heavy flavor electron production.
The red error bands represent scaling systematic uncertainties from the
dilution factor and the blue error bands represents offset systematic
uncertainties from relative luminosity and the background spin asymmetry.} 
\end{figure}

The result of the double-spin asymmetry of heavy flavor electrons
is shown in Fig.~\ref{fig:SingleEALL} and tabulated in
Table~\ref{tab:SingleEALL}.
We show systematic uncertainties for scaling and offset separately in the
figure.
The measured asymmetry is consistent with zero.

\begin{table*}[tbh]
\caption{\label{tab:SingleEALL}Data table for the $A_{LL}^{\rm{HFe}}$ result corresponding to Fig.~\ref{fig:SingleEALL}.}
\begin{ruledtabular}\begin{tabular}{c c c c c c}
$p_{T}$[GeV/$c$]&$A_{LL}^{\rm{HFe}}$& &\ \ stat. uncertainty\ \ &\ \ syst. uncertainty (offset)\ \ &\ \ syst. uncertainty (scale)\ \ \\
\hline
0.612 & 2.83$\times10^{-2}$ & \ \ \  & 2.66$\times10^{-2}$ & 0.75$\times10^{-2}$ & 0.50$\times10^{-2}$ \\
0.864 & -1.20$\times10^{-2}$ & \ \ \  & 1.21$\times10^{-2}$ & 0.30$\times10^{-2}$ & 0.08$\times10^{-2}$ \\
1.115 & 0.76$\times10^{-2}$ & \ \ \  & 1.30$\times10^{-2}$ & 0.21$\times10^{-2}$ & 0.04$\times10^{-2}$ \\
1.366 & 2.08$\times10^{-2}$ & \ \ \  & 1.63$\times10^{-2}$ & 0.18$\times10^{-2}$ & 0.10$\times10^{-2}$ \\
1.617 & -0.69$\times10^{-2}$ & \ \ \  & 2.18$\times10^{-2}$ & 0.17$\times10^{-2}$ & 0.03$\times10^{-2}$ \\
1.867 & -1.39$\times10^{-2}$ & \ \ \  & 2.68$\times10^{-2}$ & 0.16$\times10^{-2}$ & 0.03$\times10^{-2}$ \\
2.118 & 4.82$\times10^{-2}$ & \ \ \  & 3.46$\times10^{-2}$ & 0.16$\times10^{-2}$ & 0.09$\times10^{-2}$ \\
2.369 & -5.91$\times10^{-2}$ & \ \ \  & 4.40$\times10^{-2}$ & 0.16$\times10^{-2}$ & 0.11$\times10^{-2}$ \\
2.619 & -6.97$\times10^{-2}$ & \ \ \  & 5.47$\times10^{-2}$ & 0.16$\times10^{-2}$ & 0.13$\times10^{-2}$ \\
2.869 & 6.43$\times10^{-2}$ & \ \ \  & 7.07$\times10^{-2}$ & 0.16$\times10^{-2}$ & 0.12$\times10^{-2}$ \\
\end{tabular}\end{ruledtabular}
\end{table*}

 \section{\label{sec:Discussion}Discussion}

In this section, we discuss constraint of $\Delta g$ from the measured
double-spin asymmetry with an LO pQCD calculation.
In $p$$+$$p$ collisions at $\sqrt{s}=200$~GeV, heavy flavor electrons 
with momentum ranging $0.50<p_{T}<1.25$~GeV/$c$ are mainly produced by
open charm events as described in Sec.~\ref{sec:CrossSection}.
Whereas the precise mechanism for $J/\psi$ production is unknown,
unpolarized and polarized cross section of the open charm production can
be estimated with pQCD calculations.
In LO pQCD calculations, only $gg\rightarrow c\bar{c}$ and
$q\bar{q}\rightarrow c\bar{c}$ are allowed for the open charm
production.
The charm quarks are primarily created 
by the $gg$ interaction in the unpolarized hard scattering. 
In addition,
the anti-quark polarizations are known to be small from semi-inclusive
DIS measurements precisely enough that both DSSV \cite{DSSV:PRD2009}
and GRSV \cite{GRSV:PRD2001} expect
contribution of polarized $q\bar{q}$ cross section to the double-spin
asymmetry of the heavy flavor electrons in $|\eta|<0.35$ and
$p_{T}<3.0$~GeV/$c$ to be
$\sim10^{-4}$ \cite{HQSpin:arXiv2009}, which is
much smaller than the accuracy of this measurement.
Therefore,
in this analysis of $\Delta g$, we ignore the $q\bar{q}$ interaction and
assume the asymmetries are due only to the $gg$ interaction.
Under the assumption,
the spin asymmetry of the heavy flavor electrons
is expected to be
approximately proportional to the 
square of polarized gluon distribution normalized by unpolarized
distribution, $|\Delta g/g(x,\mu)|^{2}$.

We estimated the unpolarized and the polarized cross section
of charm production in $p$$+$$p$ collisions with
a LO pQCD calculation of $gg\rightarrow c\bar{c}$ \cite{LOHFXSect:PLB1994}.
For this calculation, CTEQ6M \cite{CTEQ6:JHEP2002}
was employed for the unpolarized
parton distribution functions (PDF).
For the polarized PDF, we assumed $|\Delta
g(x,\mu)|=Cg(x,\mu)$ where $C$ is a constant.
The charm quark mass was assumed as $m_{c}=$~1.4~GeV/$c^{2}$ and
the factorization scale in CTEQ6 and the renormalization scale
were assumed to be identical to
$\mu=m_{T}^{c}\equiv\sqrt{{p_{T}^{c}}^{2}+{m_{c}}^{2}}$. 

The fragmentation and decay processes were simulated with
{\sc pythia}8 \cite{PYTHIA8:ArXiv2007,PYTHIA6:ArXiv2006}.
We generated $pp\rightarrow c\bar{c} + X$ events and selected electrons
from the charmed hadrons, $D^{+}$, $D^{0}$, $D_{s}$, $\Lambda_{c}$ and
their antiparticles.
We scaled the charm quark yield in {\sc pythia} with respect to the pQCD
calculated unpolarized and polarized cross sections to obtain
unpolarized and polarized electron yields from charmed hadron decays
under these cross sections.
We also applied a pseudorapidity cut of $|\eta|<0.35$ for the
electrons
to match the acceptance of the PHENIX central arms.
The shape of the expected spin asymmetry $A_{LL}^{\rm{HFe}}(p_{T})$ is
then determined from the simulated electron yields.

\begin{figure}[tbh]
\includegraphics[width=1.0\linewidth,clip]{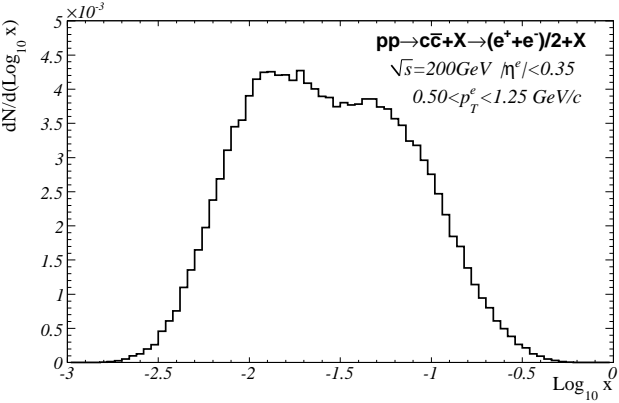}
\caption{\label{fig:XDist}(color online). Bjorken $x$ distributions of
gluons contributing the heavy flavor electron production with
momentum ranging $0.50<p_{T}<1.25$ GeV/$c$
obtained from {\sc pythia} simulation.
The distribution is normalized with respect to the number of total
generated charmed hadrons.
}
\end{figure}

Figure~\ref{fig:XDist} shows the distributions of the gluon Bjorken $x$
contributing to heavy flavor electron production
in the momentum range $0.50<p_{T}<1.25$~GeV/$c$,
from {\sc pythia}.
Using the mean and the RMS of the distribution for $0.50<p_{T}<1.25$ GeV/$c$, we
determine the mean $x$ for heavy flavor electron production to be
$\left\langle\log_{10}x\right\rangle=\xmean^{+\xrmsp}_{-\xrmsn}$.

\begin{figure}[tbh]
\subfigure[\label{fig:ALLFit:Result}Expected $A_{LL}^{\rm{HFe}}$ for
several $|\Delta g/g|$.]
{\includegraphics[width=1.0\linewidth,clip]{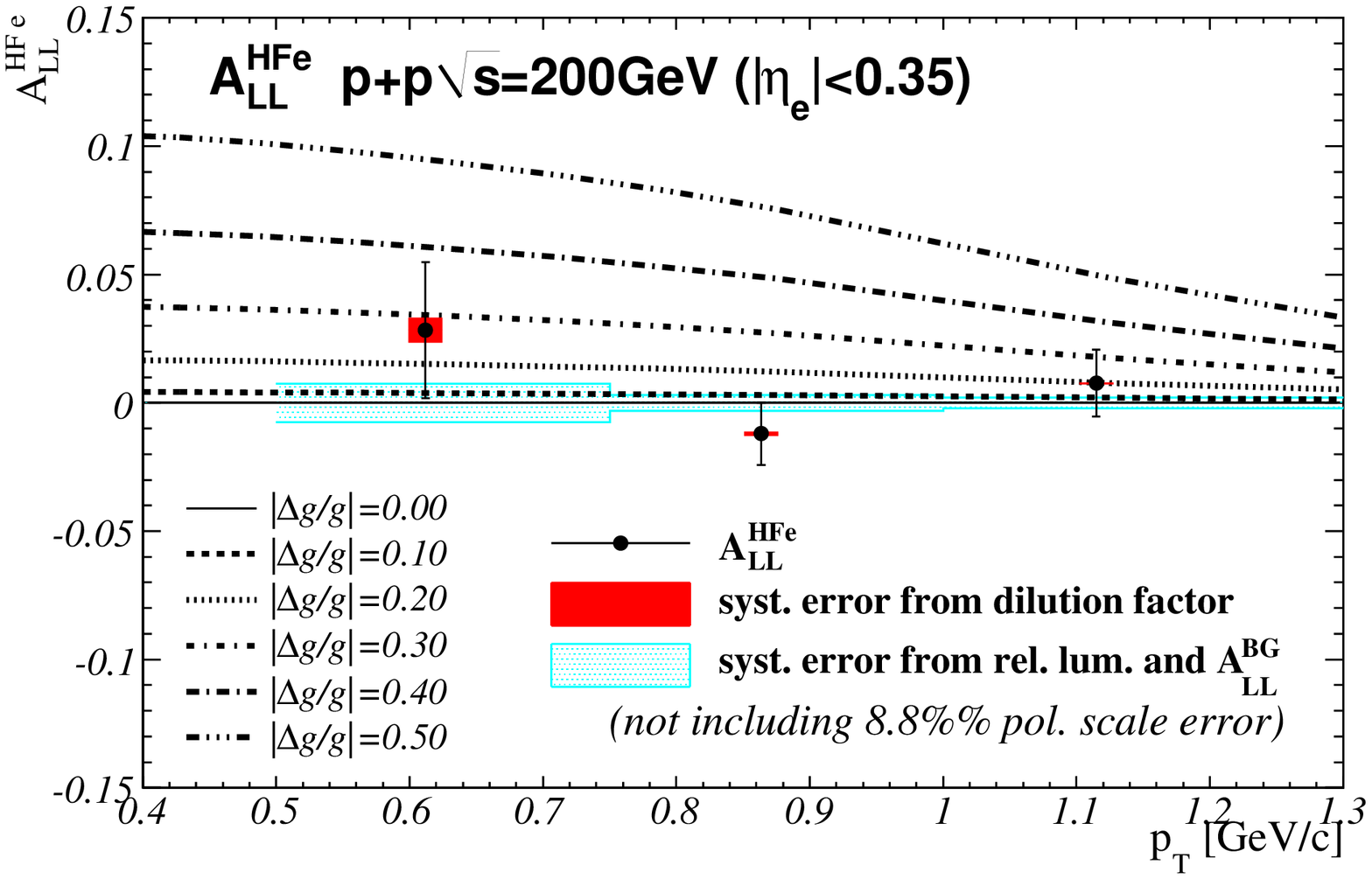}}
\subfigure[\label{fig:ALLFit:Chis}$\hat{\chi}^{2}$ curves.]{\includegraphics[width=1.0\linewidth,clip]{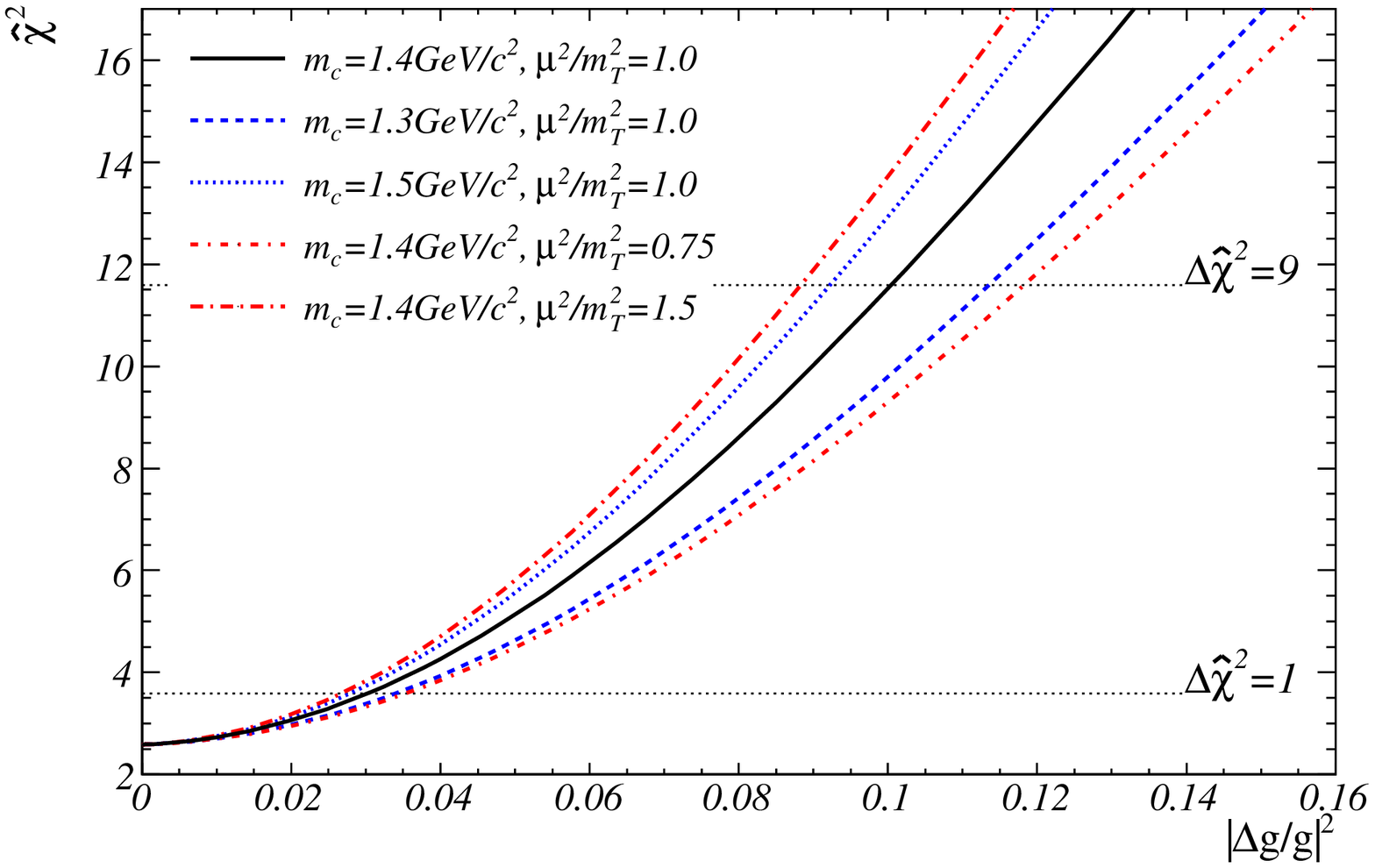}}
\caption{\label{fig:ALLFit}(color online). (a) 
$A_{LL}^{\rm{HFe}}$ for $|\Delta g/g|=0.00$, $0.10$, $0.20$,
$0.30$, $0.40$, $0.50$ are shown as the solid line, the dashed line,
the dotted line, the dashed dotted line, the long-dashed dotted line,
the dashed triplicate-dotted line respectively. They are plotted with the
measured data points and the notation for the error bars are same as
Fig.~\ref{fig:SingleEALL}. (b) $\hat{\chi}^{2}$ curves calculated 
from (a) as a function of $|\Delta g/g|^{2}$. The black solid line
is the default configuration. The blue curves are
after changing the charm mass to 1.3 GeV/$c^{2}$ (dashed line) and
to 1.5 GeV/$c^{2}$ (doted line) and the red curves are after
changing the scale $\mu^{2}$ to $0.75{m_{T}^{c}}^{2}$ (dashed
dotted line) and
$1.5{m_{T}^{c}}^{2}$ (long-dashed dotted line).}
\end{figure}

We calculated expected $A_{LL}^{\rm{HFe}}(p_{T})$
by varying $C=|\Delta g/g|$.
Figure~\ref{fig:ALLFit:Result} shows several of these curves, along with
the measured points.
$\chi^{2}$ values are calculated for each value of $C$, along with
related uncertainties.
By assuming that the systematic uncertainties on the points
are correlated and represent global shifts,
we defined the quantity $\hat{\chi^{2}}$ as
\begin{equation}
 \label{eq:EffectiveChis}
\begin{array}{rl}
 \hat{\chi}^{2}(C)&\equiv-2\log\left(\left(2\pi\right)^{\frac{n}{2}}\hat{P}(C)\right)
  \nonumber \\
 \hat{P}(C)&\equiv\int dpdqN(p)N(q)\times \nonumber \\
 &\prod_{i=1}^{n}N\left(\frac{\left(y_i+p\epsilon_{\rm{syst}}^{i\
				      \rm{offset}}-\left(1+q\gamma_{\rm{syst}}^{i\ \rm{scale}}\right)f(x_i;C)\right)}{\epsilon_{\rm{stat}}^{i}}\right) \\
 &\gamma_{\rm{syst}}^{i\ \rm{scale}}=\sqrt{\left(\frac{\epsilon_{\rm{syst}}^{i\ \rm{scale}}}{y_i}\right)^{2}+\left(\frac{\Delta(P_BP_Y)}{P_BP_Y}\right)^{2}}, \nonumber \\
\end{array}
\end{equation}
where $N(X)$ denotes normal probability distribution,
i.e. $N(X)=1/\sqrt{2\pi}\exp(-X^{2}/2)$, $n$ is the number
of the data points and equal to three, and for the $i$-th data point,
$x_i$ is the $p_{T}$ value, $y_i$ is the $A_{LL}$ value, and
$\epsilon_{\rm{stat}}^{i}$, 
$\epsilon_{\rm{syst}}^{i\ \rm{offset}}$ and $\epsilon_{\rm{syst}}^{i\
\rm{scale}}$ represent the statistical, offset systematic and 
scaling systematic uncertainties, respectively.
$f(p_{T};C)$ denotes the expected $A_{LL}(p_{T})$ for the
parameter of $C=|\Delta g/g|$.
$\Delta(P_BP_Y)$ is an uncertainty for polarization mentioned in
Sec.~\ref{sec:HelicityAsymmetry}.
%
If we set the systematic uncertainties,
$\epsilon_{\rm{syst}}^{\rm{offset}}$ and $\gamma_{\rm{syst}}^{i\
\rm{scale}}$, to zero, the newly defined $\hat{\chi^{2}}$ is consistent
with the conventional $\chi^{2}$.

The resulting $\hat{\chi}^{2}$ curve is shown in
Fig.~\ref{fig:ALLFit:Chis}, plotted as a function of $C^{2}=|\Delta 
g/g|^{2}$ because the curvature becomes almost parabolic.
The minimum of $\hat{\chi}^{2}$, $\hat{\chi}^{2}_{{\rm min}}$, is located
at $|\Delta g/g|^{2}=0.0$ which is the boundary of $|\Delta g/g|^{2}$.
$\Delta\hat{\chi}^{2}\equiv\hat{\chi}^{2}-\hat{\chi}^{2}_{{\rm min}}=1$
and $9$ were
utilized to determine $1\sigma$ and $3\sigma$ uncertainties.
With these criteria, we found the constraints on the gluon
polarization are
$|\Delta g/g(\left\langle\log_{10}x\right\rangle,\mu)|^{2}< 
 \dggsqrerrlim\times10^{-2}$($1\sigma$) and
 $\dggsqrthreeerrlim\times10^{-2}$($3\sigma$).
The constraints are consistent with theoretical expectations for
$\Delta g/g(x,\mu)$ at
$\left\langle\log_{10}x\right\rangle=\xmean^{+\xrmsp}_{-\xrmsn}$ and 
$\mu=1.4$~GeV which are
$\sim\dggdssv$ from DSSV,
$\sim\dgggrsvs$ from GRSV(std) and
$\sim\dgggrsvv$ from GRSV(val)
using CTEQ6 for the unpolarized PDF.

The effects of the charm quark mass and scale factor in the
cross section calculation were also checked by varying the charm mass from
$m_{c}=1.3$~GeV/$c^{2}$ to 1.5~GeV/$c^{2}$ and the scale 
to $\mu^{2}=0.75{m_{T}^{c}}^{2}$ and $1.5{m_{T}^{c}}^{2}$.
Figure~\ref{fig:ALLFit:Chis} also shows the resulting $\hat{\chi}^{2}$ curves.
Considering the variation of the crossing position at
$\Delta\hat{\chi}^{2}=1$,
the constraint including the uncertainties from the charm mass and the
scale can be represented as $|\Delta
g/g|^{2}<(\dggsqrerrlim_{-\diffdggsqrerrmassn}^{+\diffdggsqrerrmassp}{\rm{(mass)}}_{-\diffdggsqrerrscalen}^{+\diffdggsqrerrscalep}{\rm{(scale)}})\times10^{-2}$($1\sigma$).

The integral of the CTEQ6 unpolarized PDF in the sensitive $x$
region of
$\left\langle\log_{10}x\right\rangle=\xmean^{+\xrmsp}_{-\xrmsn}$ and
$\mu=1.4$~GeV is $\int_{0.01}^{0.08}dxg(x,\mu)=4.9$.
Hence the constraint on the integral of the polarized PDF at $1\sigma$
corresponds to
$|\int_{0.01}^{0.08}dx\Delta g(x,\mu)|<0.85$.
This study also highlights the possibility for constraining $\Delta g$
in this Bjorken $x$ region more precisely in the future
with higher statistics and
higher beam polarizations.
 \section{\label{sec:Summary}Summary}

We have presented a new analysis method for identifying heavy flavor electrons at
PHENIX.
With this new method, the signal purity is improved by a factor of
about 1.5 around $0.75\simlt p_{T} \simlt 2.00$~GeV/$c$ due to the
rejection of photonic electrons by the HBD.
We have reported on the first measurement of the longitudinal
double-spin asymmetry of heavy flavor electrons, which are consistent
with zero. 
Using this result, we estimate a 
constraint of $|\Delta g/g
(\log_{10}x=\xmean^{+\xrmsp}_{-\xrmsn},\mu=m_{T}^{c})|^{2}<\dggsqrerrlim\times10^{-2}$($1\sigma$). 
This value is consistent with the existing theoretical expectations with
GRSV and DSSV.
With improved statistics and polarization, the helicity
asymmetry of heavy flavor electron production can provide more
significant constraints on the gluon polarization, and complement other
measurements of $\Delta G$.

\section*{ACKNOWLEDGMENTS}

We thank the staff of the Collider-Accelerator and Physics
Departments at Brookhaven National Laboratory and the staff of
the other PHENIX participating institutions for their vital
contributions.  
We thank Marco Stratmann for detailed discussions about constraining the
gluon polarization and for the preparation of codes used to calculate the
cross sections.
We acknowledge support from the 
Office of Nuclear Physics in the
Office of Science of the Department of Energy, the
National Science Foundation, Abilene Christian University
Research Council, Research Foundation of SUNY, and Dean of the
College of Arts and Sciences, Vanderbilt University (U.S.A),
Ministry of Education, Culture, Sports, Science, and Technology
and the Japan Society for the Promotion of Science (Japan),
Conselho Nacional de Desenvolvimento Cient\'{\i}fico e
Tecnol{\'o}gico and Funda\c c{\~a}o de Amparo {\`a} Pesquisa do
Estado de S{\~a}o Paulo (Brazil),
Natural Science Foundation of China (P.~R.~China),
Ministry of Education, Youth and Sports (Czech Republic),
Centre National de la Recherche Scientifique, Commissariat
{\`a} l'{\'E}nergie Atomique, and Institut National de Physique
Nucl{\'e}aire et de Physique des Particules (France),
Bundesministerium f\"ur Bildung und Forschung, Deutscher
Akademischer Austausch Dienst, and Alexander von Humboldt Stiftung (Germany),
Hungarian National Science Fund, OTKA (Hungary), 
Department of Atomic Energy and Department of Science and Technology (India), 
Israel Science Foundation (Israel), 
National Research Foundation and WCU program of the 
Ministry Education Science and Technology (Korea),
Ministry of Education and Science, Russian Academy of Sciences,
Federal Agency of Atomic Energy (Russia),
VR and Wallenberg Foundation (Sweden), 
the U.S. Civilian Research and Development Foundation for the
Independent States of the Former Soviet Union, 
the Hungarian American Enterprise Scholarship Fund,
and the US-Israel Binational Science Foundation.
 

\end{document}